\documentclass{aa}
\usepackage{txfonts}
\usepackage[dvips]{graphicx}
\usepackage{amsfonts}
\usepackage{amssymb}
\usepackage{multirow}
\usepackage{natbib}
\bibpunct{(}{)}{;}{a}{}{,}

\newcommand{\te}{$T_{\rm eff}$}
\newcommand{\logg}{$\log{g}$}
\newcommand{\vsini}{$v\sin{i}$}
\newcommand{\kms}{km\,s$^{-1}$}
\newcommand{\GD}{$\gamma$~Dor}
\newcommand{\DSct}{$\delta$~Sct}
\newcommand{\cd}{d$^{-1}$}
\newcommand{\mhz}{$\mu$Hz}

\begin{document}
\title{Denoising spectroscopic data by means of the improved Least-Squares Deconvolution method\thanks{Based on the data gathered with the {\sc HERMES} spectrograph, installed at the Mercator
    Telescope, operated on the island of La Palma by the Flemish Community, at
    the Spanish Observatorio del Roque de los Muchachos of the Instituto de
    Astrof\'{\i}sica de Canarias and supported by the Fund for Scientific
    Research of Flanders (FWO), Belgium, the Research Council of K.U.Leuven,
    Belgium, the Fonds National de la Recherche Scientific (F.R.S.--FNRS),
    Belgium, the Royal Observatory of Belgium, the Observatoire de Gen\`eve,
    Switzerland and the Th\"uringer Landessternwarte Tautenburg, Germany.}$^,$\thanks{Based on the data extracted from the ELODIE archive and the ESO Science Archive Facility under request number TVanReeth63233}$^,$\thanks{The software presented in this work is available upon request from: Andrew.Tkachenko@ster.kuleuven.be}}
\author{A. Tkachenko\inst{1}$^,$\thanks{Postdoctoral Fellow of the Fund for Scientific Research (FWO), Flanders, Belgium} \and T. Van Reeth\inst{1} \and V. Tsymbal\inst{2} \and
C. Aerts\inst{1,3} \and O. Kochukhov\inst{4} \and J.
Debosscher\inst{1}} \institute{Instituut voor Sterrenkunde, KU
Leuven, Celestijnenlaan 200D, B-3001 Leuven, Belgium\\
\email{Andrew.Tkachenko@ster.kuleuven.be} \and Tavrian National
University, Department of Astronomy, Simferopol, Ukraine \and
Department of Astrophysics, IMAPP, University of Nijmegen, PO Box
9010, 6500 GL Nijmegen, The Netherlands \and Department Physics and
Astronomy, Uppsala University, Box 516, 751 20, Uppsala, Sweden}
\date{Received date; accepted date}
\abstract{The MOST, CoRoT, and $Kepler$ space missions led to the
discovery of a large number of intriguing, and in some cases unique,
objects among which are pulsating stars, stars hosting exoplanets,
binaries, etc. Although the space missions deliver photometric data
of unprecedented quality, these data are lacking any spectral
information and we are still in need of ground-based spectroscopic
and/or multicolour photometric follow-up observations for a solid
interpretation.}{Both faintness of most of the observed stars and
the required high signal-to-noise ratio (S/N) of spectroscopic data
imply the need of using large telescopes, access to which is
limited. In this paper, we look for an alternative, and aim for the
development of a technique allowing to denoise the originally low
S/N (typically, below 80) spectroscopic data, making observations of
faint targets with small telescopes possible and effective.}{We
present a generalization of the original Least-Squares Deconvolution
(LSD) method by implementing a multicomponent average profile and a
line strengths correction algorithm. We test the method both on
simulated and real spectra of single and binary stars, among which
are two intrinsically variable objects.}{The method was successfully
tested on the high-resolution spectra of Vega and a $Kepler$ star,
KIC04749989. Application to the two pulsating stars, 20~Cvn and
HD\,189631, showed that the technique is also applicable to
intrinsically variable stars: the results of frequency analysis and
mode identification from the LSD model spectra for both objects are
in good agreement with the findings from literature. Depending on
S/N of the original data and spectral characteristic of a star, the
gain in S/N in the LSD model spectrum typically ranges from 5 to 15
times.}{The technique introduced in this paper allows an effective
denoising of the originally low S/N spectroscopic data. The high S/N
spectra obtained this way can be used to determine fundamental
parameters and chemical composition of the stars. The restored LSD
model spectra contain all the information on line profile variations
present in the original spectra of, e.g., pulsating stars. The
method is applicable to both high- ($>$30\,000) and low-
($<$30\,000) resolution spectra, though the information that can be
extracted from the latter is limited by the resolving power itself.}
 \maketitle

\section{Introduction}

The recent launches of the space missions like MOST
\citep{Walker2003}, CoRoT \citep{Auvergne2009} and $Kepler$
\citep{Gilliland2010}, led to the discovery of numerous pulsating
stars. There is a wealth of information that can be extracted from
nearly continuous micro-magnitude precision photometric
observations, but additional ground-based observations are essential
for an accurate analysis and correct interpretation of the observed
light variability. Ground based spectroscopic observations are
necessary to determine the fundamental parameters like effective
temperature \te\ and surface gravity \logg. This allows to
discriminate between, e.g., Slowly Pulsating B
\citep[SPB,][]{Waelkens1991} and $\gamma$\,Dor
\citep{Cousins1992,Krisciunas1993,Balona1994} variable stars which
show the same type of variability in their light curves but are
located in different regions of the Hertzsprung-Russell (HR)
diagram. Moreover, measuring the projected rotational velocity
\vsini\ from broadening of spectral lines helps to constrain the
true rotation rate of the star, which in turn helps to interpret the
observed characteristic frequency patterns. Spectroscopic
measurements allow to study binarity while detailed abundance
analysis of high-resolution spectra might help to unravel chemical
peculiarities in stellar atmospheres. This kind of analysis usually
requires spectroscopic data of high signal-to-noise ratio (S/N)
which is hard to achieve for faint stars ($<$ 12 mag) observed by
the satellites, unless one has an easy access to 4-m class
telescopes. However, smaller class telescopes can successfully
compete with the bigger ones if a technique allowing for a serious
increase of S/N in stellar spectra without loosing important
information (like in the case of binning or smoothing the spectra)
exists.

Least-squares deconvolution (LSD) is a powerful tool for extraction
of high quality, high S/N average line profiles from stellar
spectra. The technique was first introduced by \citet{Donati1997}
and is based on two fundamental assumptions: (i) all spectral lines
in the stellar spectrum are similar in shape, i.e. can be
represented by the same average profile scaled in depth by a certain
factor; (ii) the intensities of overlapping spectral lines add up
linearly. When applied to unpolarized spectra, this method has some
similarities to the broadening function technique introduced by
\citet{Rucinski1992,Rucinski2002} and is different in that it
represents a convolution of the unknown average profile with the
line mask that contains information about the position of spectral
lines and their strengths. Unlike the broadening function technique
that uses unbroadened synthetic or observed spectrum as a template,
the LSD method does not require a template spectrum which makes it a
more versatile and less model dependent technique. The technique is
used to investigate the physical processes taking place in stellar
atmospheres and affecting all spectral line profiles in a similar
way. This includes the study of line profile variations (LPV) caused
by, e.g., orbital motion of the star and/or stellar surface
inhomogeneities \citep[see e.g.,][]{Lister1999,Jarvinen2010}.
However, its widest application nowadays is the detection of weak
magnetic fields in stars over the entire HR diagram based on
Stokes~$V$ (circular polarization) observations \citep[see
e.g.,][]{Shorlin2001,Donati2008,Alecian2011,Kochukhov2011,Silvester2012,Aerts2013}.

There have been several attempts to go beyond using the LSD
technique as an LPV and/or magnetic field detection tool.
\citet{Sennhauser2009,Sennhauser2010} attempted to generalize the
LSD method by overcoming one of its fundamental assumptions, namely
a linear addition of contributions from different lines. In their
Nonlinear Deconvolution with Deblending (NDD) approach, the authors
assume each single line profile to be approximated according to the
interpolation formula given by \citet{Minnaert1935} and strong,
optically thick lines to add nonlinearly. \citet{Donati2003} and
\citet{Folsom2008}, in their studies of magnetic and chemical
surface maps, respectively, treat the ``classical'' LSD profile as a
real, isolated spectral line with average properties of all
concerned contributions in a given wavelength range. According to
\cite{Kochukhov2010}, such an approach must be applied with caution
as it is justified in only a limited parameter range. In particular,
the authors find that Stokes $I$ (intensity) and Stokes $Q$ (linear
polarization) LSD-profiles do not resemble a real spectral line with
average parameters, whereas Stokes $V$ (circular polarization)
LSD-profile behaves very similar to a properly chosen isolated
spectral line for the magnetic fields weaker than 1 kG.  In their
work, \cite{Kochukhov2010} also present a generalization of the
standard LSD technique implemented in the newly developed
\emph{improved least-squares deconvolution} (iLSD) code which, among
other things, allows for the calculation of the so-called
multiprofile LSD. The latter assumes that each spectral line in the
observed spectrum is represented by a superposition of $N$ different
scaled LSD profiles and shall have a practical application for the
stellar systems consisting of two and more stellar components.

In this paper, we present a further generalization of the LSD
method, by considering individual line strengths correction. We do
not focus on the analysis of the average profiles themselves, but
rather consider them to be an intermediate step towards building
high S/N model spectra from the originally low quality spectroscopic
observations. As such, we propose a new application of the LSD
method, namely an effective ``denoising'' of the spectroscopic data.
In Section~2, we give a mathematical description of the LSD method
following the pioneering work by \citet{Donati1997}, and propose an
alternative way of computing average profiles by solving an inverse
problem. Section~3 discusses the implementation of our
generalization of the standard method, which includes both the
calculation of the multicomponent LSD profile and a line strengths
correction algorithm. We refer to Section~4 for the application of
the method to both simulated and real, observed spectra of single
stars, whereas Section~5 explores the possibility of applying the
technique to binary star systems. We close the paper with Section~6,
where a discussion and the conclusions are presented.

\section{Least-squares deconvolution: theoretical
background}\label{Section2}

The LSD technique allows to compute a mean profile representative of
all individual spectral lines in a particular wavelength range and
which is formally characterized by extremely high S/N. Within the
fundamental assumptions, the problem is formulated mathematically as
a convolution of an unknown mean profile and an a priori known line
mask:

\begin{equation}
I = M \ast Z(\upsilon).
\end{equation}
Alternatively, the expression can be represented as a simple matrix
multiplication:

\begin{equation}
\mathbf{I} = \mathbf{M} \cdot \mathbf{Z,}
\end{equation}
with $\mathbf{I}$ an $i$-element model spectrum, $\mathbf{M}$ an $i
\times j$-element matrix that contains information about the
position of the lines and their central depths, and $\mathbf{Z}$ a
$j$-element vector representing mean for all individual spectral
lines profile. We refer to \citet{Donati1997} and
\citet{Kochukhov2010} for a comprehensive theoretical description of
the method.

As already mentioned, the line mask $\mathbf{M}$ contains line
positions and their relative strengths and can be compiled from any
atomic lines database \citep[e.g., Vienna Atomic Lines Database,
VALD,][]{Kupka1999} or pre-calculated using a spectral synthesis
code. One of the technique's fundamental assumptions requires
hydrogen and helium as well as the metal lines exhibiting strong,
damping wings to be excluded from the mask, in the case of hot and
cool stars, respectively.

In practice, the extraction of a mean profile directly from the
observations is an inverse problem, which we solve by means of the
Levenberg- Marquardt algorithm \citep{Levenberg1944,Marquardt1963}.
We do not make any a priory assumptions neither on the shape nor on
the depth of the LSD-profile but start the calculations from the
constant intensity value. The profile is computed in each point of
the specified velocity grid. To speed up the calculations, we use a
modified, fast version of the Levenberg-Marquardt algorithm
developed by \cite{Piskunov2002}, which is widely used for the
reconstruction of the temperature, abundance, and magnetic field
maps of stellar surfaces
\citep[e.g.,][]{Luftinger2010,Nesvacil2012,Kochukhov2013}. The
essential difference of the modified algorithm from the original one
is that it additionally adjusts the value of the damping parameter
within each iteration step, keeping the Hessian matrix unchanged.
This significantly speeds up the calculations without affecting the
convergence of the method.

\section{Improvements of the technique}\label{Section3}

In this section, we present our generalization of the original LSD
method by implementing a multiprofile LSD and a line strengths
correction algorithm.

\subsection{Multiprofile LSD}

\begin{figure}[t]
\includegraphics[scale=0.92]{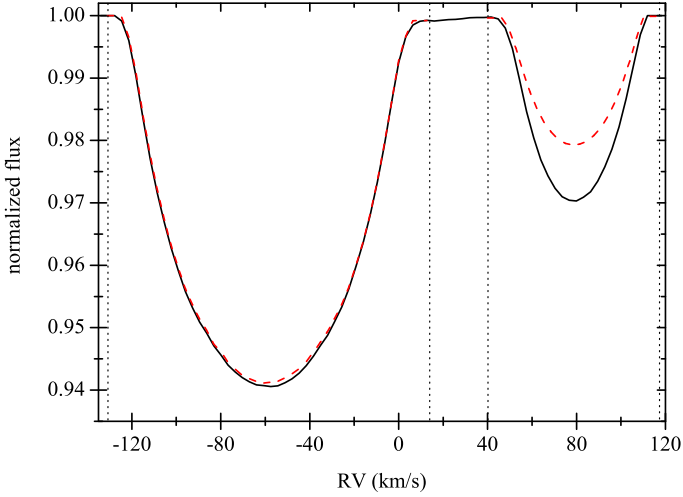}
\caption{{\small LSD-profile(s) of a synthetic binary system
computed assuming standard (solid, black line) and multiprofile
(dashed, red line) techniques. The vertical dotted lines indicate
the velocity ranges for the two stars.}} \label{Figure1}
\end{figure}

As already mentioned, the multiprofile LSD approach assumes that
each line in the observed spectrum is represented by a superposition
of N different mean profiles. According to \citet{Kochukhov2010}, in
that case one can still use the original matrix formulation of the
LSD technique if one assumes a composite line pattern matrix of
$i\times(j\cdot N)$ elements, with $N$ the number of different mean
profiles. In the case of our iterative method, we simultaneously
solve for $N$ different LSD-profiles spanning certain velocity
ranges and characterized by different line masks which allow for
diversity in atmospheric conditions. This has a practical
application for binary and multiple star systems where individual
stellar components can have quite different atmospheric parameters
and represent their own sets of spectral lines formed under
different physical conditions.

The original LSD technique introduced by \citet{Donati1997} also
allows to deal with binary star systems but only with those composed
of two identical stars. However, if two stars that form a binary
system have different atmospheric conditions (effective temperature,
surface gravity, chemical composition, etc.), an assumption of the
standard LSD technique about common line mask for both stellar
components appears to be incorrect. In this case, the standard
method provides a potential of detecting (possibly weak)
contribution of the companion star in the composite spectrum of a
binary, but the depth of the (composite) average profile is largely
influenced by the assumed line mask. This is illustrated in
Figure~\ref{Figure1} that compares LSD-profiles of a synthetic
binary system computed assuming the standard technique as formulated
by \citet{Donati1997} and a line mask of the hotter primary common
for both stars (solid line), and the multiprofile technique with the
difference in atmospheric conditions of the two stars taken into
account (dashed lines). This difference is accounted for by using
masks computed assuming different fundamental parameters for the two
stars. In Figure~\ref{Figure1}, we assume
\te$^{1,2}$=8\,500/7\,300~K, \logg$^{1,2}$=3.4/3.5~dex,
[M/H]$^{1,2}$=\ \ 0.0/-0.3~dex, \vsini$_{1,2}$=60/30~\kms\ for the
primary and secondary, respectively, and fix microturbulent velocity
to 2~\kms\ for both stars. The average profiles are computed at
phase of large RV separation assuming non-overlapping velocity grids
for the two stars (indicated by the vertical dotted lines in
Figure~\ref{Figure1}). Apparently, we obtain an overestimated
contribution of the (cooler) secondary component when relying on the
standard technique. When scaled to unity, the profiles obtained from
different techniques have the same shape and deliver RVs which
differ from each other by less than 100~m\,s$^{-1}$. In Section~5,
we will come back to binary stars where, in particular, a discussion
on applicability of the multiprofile LSD technique to different
orbital phases will be discussed.

\begin{figure}[t]
\includegraphics[scale=0.365]{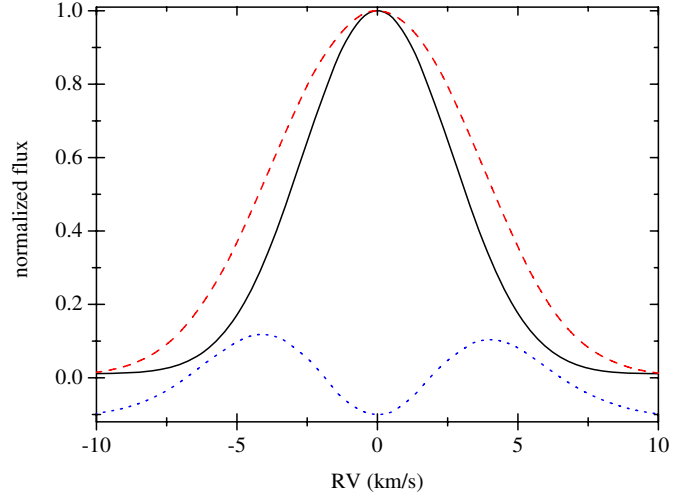}
\caption{{\small LSD-profiles computed for two different groups of
spectral lines: with relative strengths $<0.4$ (solid, black line)
and $>0.4$ (dashed, red line). The difference between the two
profiles (dotted, blue line) is shown at the bottom and was shifted
in Y by value of -0.1 for clarity.}} \label{Figure2}
\end{figure}

The multiprofile LSD technique (to some extent) also allows to
overcome one of the fundamental assumptions stating that all
spectral lines in the observed spectrum are similar in shape. This
assumption is justified for the majority of metallic lines in the
spectrum of rapidly rotating stars as the rotational broadening
clearly dominates over the other sources of broadening in this case.
However, it fails for the slow rotators. Figure~\ref{Figure2} shows
normalized LSD-profiles computed for two different groups of
spectral lines selected based on their theoretical relative
strengths. The two profiles clearly have different shapes which can
also be seen from their difference shifted to the bottom of the
figure for clarity. When implementing this generalization of the
standard LSD technique, one has to remember that the strengths of
spectral lines in the spectrum are smoothly distributed from 0 to 1.
Representation of the spectrum by a convolution of, e.g., two mean
profiles with lines from the mask, where all ``weak'' and ``strong''
(below and above certain line strengths limit) lines are represented
by their own LSD profile, assumes rather a step-like distribution of
the line strengths. To overcome the problem of this strict division
into groups with sharp edges, when solving simultaneously for $n$
mean profiles, we represent each line in the observed spectrum as a
convolution of the corresponding entry from the line mask and the
LSD profile computed by means of a parabolic interpolation between
the $n$ mean profiles we are solving for.

Our generalization towards a multicomponent LSD profile implies that
one is free to compute $N\cdot n$ average profiles at the same time,
with $N$ the number of individual stellar components (two for a
binary, three for a triple system, etc.) and $n$ the number of
LSD-profiles per stellar component. In the calculation of the $n$
average profiles, the line weights are assigned based entirely on
their relative strengths information from the mask. This implies
that a pair of lines that have similar intrinsic strengths but,
e.g., significantly different pressure broadening will be given
(nearly) equal weights. This in turn means that besides hydrogen and
helium lines, the metal lines with broad, damping wings should also
be excluded from the calculations, regardless of their intrinsic
strength.

Our experience shows that for moderately to rapidly rotating stars
(\vsini$>$30~\kms), with the rotation being dominant source of
spectral line broadening, calculation of $n=2$ average profiles
seems to be an optimal choice. For slowly rotating stars,
calculation of the additional third LSD-profile is required to
better account for the shape difference between the lines assigned
to different groups. Solving for $n>3$ profiles is also possible
but, in most cases, does not seem to be really feasible. In this
case, the lines from the two groups representing the weakest
spectral lines show negligible difference in the intrinsic shapes,
and the reduced number of lines per group implies that at least one
of the (probably weakest) LSD-profiles will be badly defined.

\subsection{Line strengths correction algorithm}

\begin{figure*}[t]
\includegraphics[scale=0.9]{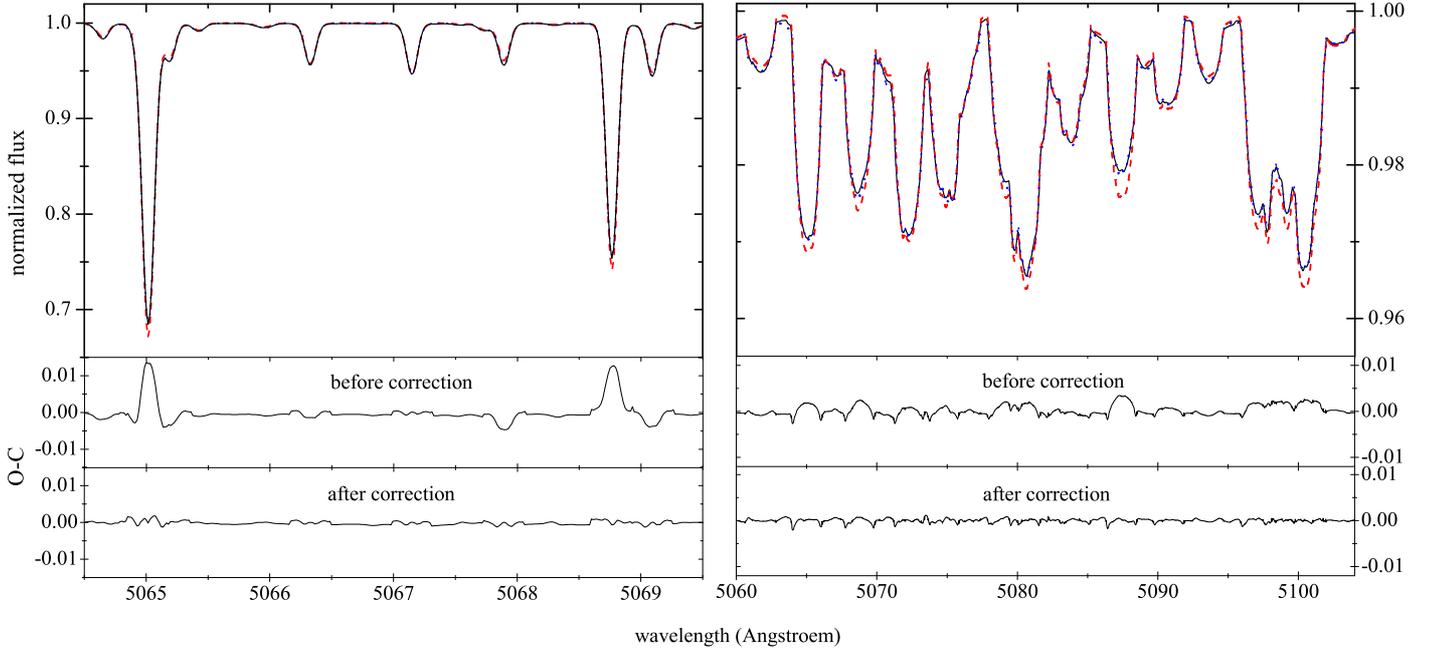}
\caption{{\small Comparison of the synthetic spectrum (solid, black
line) with the LSD models before (dashed, red line) and after
(dotted, blue line) line strengths correction. The corresponding
``O-C residuals'' (synthetic minus LSD-model spectrum) are shown in
the middle and bottom panels.} Left and right panels represent
different rotation rates: \vsini=2~\kms\ and 60~\kms.}
\label{Figure3}
\end{figure*}

Although a convolution of $n$ mean profiles (see previous Section
for definition of $n$) with lines from the mask gives a better
representation of the observed spectrum than the standard technique,
it is still not as good as one would like to have. The reason is
obviously the second fundamental assumption of linear addition of
overlapping spectral lines. As pointed out by \citet{Kochukhov2010}
and has been attempted by \citet{Reiners2003} in their ``Physical
Least-Squares Deconvolution'' (PLSD), one way to overcome this
limitation would be to solve simultaneously for the mean profile and
individual line strengths. \citet{Sennhauser2009}, on the other
hand, suggest to use a non-linear line addition law of the following
form:
\begin{equation}
R_{tot} = R_1+R_2-R_1\cdot R_2, \label{Eq.: non-linear_addition}
\end{equation}
where $R_{tot}$ stands for the depth of a blend formed of two
individual lines with depths $R_1$ and $R_2$. The line depth $R$ is
related to the line strength $r$ (to which we will refer in the
subsequent sections) via the simple equation $R=1-r$. As pointed out
by \citet[][Figure~4 and description therein]{Kochukhov2010} and
verified by us, the proposed non-linear law provides a good
representation of the naturally blended spectral lines with
overlapping absorption coefficients. However, the residual
intensities of the lines that are blended due to the limited
resolving power of the instrument or due to the stellar rotation
indeed add up linearly, just as the LSD technique assumes. In this
case, the non-linear law proposed by \citet{Sennhauser2009} often
underestimates the depth of a linearly blended line and thus appears
to be inferior to the LSD description of the spectrum. The idea of
solving simultaneously for the mean profile and individual line
strengths would certainly be a step in the right direction but this
problem is highly ill-posed and one could hardly be able to come to
a unique solution. We thus decided to separate the two problems,
calculation of the LSD profile and a correction of the individual
line strengths, and do it iteratively until the results of both
applications do not change anymore. The individual contributions are
optimized ``line by line'' which means that only one, single line is
optimized at a given instant of time, but taking into account
contributions from neighbouring lines when computing the model
spectrum. An exception is made for the lines with high percentage
overlapping absorption coefficients, i.e. the lines which naturally
locate too close to each other to be resolved by a given instrument.
In other words, we define the lines as unresolved when they
(naturally) locate to each other closer than it is assumed by the
step width of the velocity grid on which a LSD-profile is computed.
The RV step width depends on the resolving power of the instrument,
and is determined directly from the observations. In this case,
instead of optimizing individual line strengths, we solve for their
sum accounting for theoretical percentage contribution of each
individual line into the sum. Given a large number of lines we
consider in our calculations, it can also happen that, according to
our definition, arbitrary lines 1--2 and 2--3 are not resolved but
the lines 1--3 are resolved. Moreover, a series of lines can be much
longer than that and may include up to a couple of dozens of
members. Our tests show that in such a case, solving for the sum of
all the contributions becomes a very degenerate problem with
unstable solution. Thus, in our approach, in the above mentioned
three lines situation, we optimize the sum of strengths for lines 1
and 2, and then proceed with the third line, ignoring the fact that
the 2--3 pair is unresolved. One can also do contrary, i.e. solve
for the sum for the 2--3 pair and then proceed with the line 1, but
though a certain (small) difference between the two cases can be
observed during the first a couple of iterations, the end result is
found to be same. During the line strengths correction, the mean
profiles are normalized to unity to prevent a drifting of both the
profiles and individual line strengths. The optimization algorithm
can handle both single- and multi-profile LSD. In the latter case, a
spectral line in the model spectrum is represented by a convolution
of the corresponding entry from the mask with the average profile
computed by means of (parabolic) interpolation between the $n$
components of the multiprofile LSD (see previous section for the
definition of $n$). After the line strengths correction, the
(multicomponent) LSD profile is recomputed based on the improved
line mask, and the whole cycle is repeated until no significant
changes in the profiles neither in the individual line strengths
occur anymore. The efficiency of our approach will be tested and
discussed in the next sections.

We use a golden section search algorithm to minimize an O--C
(``Observed''--``Calculated'') function in the wavelength range
determined by the shape of the LSD profile and a position of the
line in question. The golden section search is a technique for
finding the extremum by successively narrowing the range of values
inside which the extremum is known to exist.

Our experience shows that two ``global'' iterations, each of which
includes the calculation of the LSD profile and typically a dozen of
iterations for correction of the individual line strengths, is
sufficient. Figure~\ref{Figure3} illustrates the results of the line
strengths correction by comparing the initial spectrum to the two
LSD models computed before and after the correction. The synthetic
spectrum has been computed for \te=8\,500~K, \logg=3.4~dex,
microturbulence $\xi$=2.0~\kms, and solar chemical composition. We
consider two cases with very different rotation rates, characterized
by a \vsini\ of 2~\kms\ (top) and 60~\kms\ (bottom). In both cases,
the initial LSD model badly describes the intensity spectrum of the
star while the match is nearly perfect after application of the line
strengths correction algorithm.

\section{Application to single stars spectra}

In this section, we test our improved LSD technique on spectra of an
artificial star, verify the influence of different line masks on the
final results, and compare our method to the simple ``running
average'' smoothing algorithm. The method is then applied to the
spectra of the ``standard'' star Vega, KIC04749989, a \GD-\DSct\
hybrid pulsator in the $Kepler$ field, and to the spectra of 20~CVn
and HD\,189631, a mono-periodic \DSct- and a multi-periodic \GD-type
pulsating star, respectively.

\subsection{Simulated data}

To test the influence of the initial model on the final results, we
computed a certain number of line masks assuming different
atmospheric parameters. We separately tested the impact of \te,
\logg, and [M/H], while the microturbulence was fixed to 2~\kms. The
effect of different noise level in the ``observed'' spectrum is also
investigated. In this section, we refer to the synthetic spectrum of
an artificial star as to the ``observed'' spectrum for convenience.

\begin{figure}[t]
\includegraphics[scale=0.83]{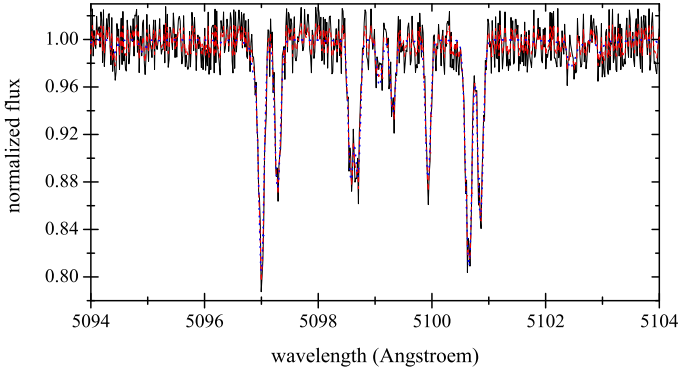}
\caption{{\small ``Observed'' spectra of an artificial star assuming
different value of S/N: 35 (solid, black line), 70 (dashed, red
line), and infinity (dotted, blue line)}} \label{Figure4}
\end{figure}

Both, synthetic spectra and individual line strengths for the masks
have been computed with the SynthV code \citep{Tsymbal1996} based on
the atmosphere models computed with the most recent version of the
LLmodels code \citep{Shulyak2004}. The information about positions
of the individual lines was extracted from the VALD database
\citep{Kupka1999}. We assume the star to have the same atmospheric
parameters as in the previous section, i.e. \te=8\,500~K,
\logg=3.4~dex, $\xi$=2.0~\kms, \vsini=2.0~\kms, and solar chemical
composition. To simulate the data of different S/N, we added
Gaussian white noise to our spectra. Figure~\ref{Figure4}
illustrates a part of the ``observed'' spectrum assuming three
different values of S/N.

The whole idea of the experiment presented here is as follows: we
start with the initial ``observed'' spectrum, of which typically
800--1000~\AA\ wide wavelength range free of Balmer lines is used to
compute a multicomponent LSD profile as described in Section~3.1.
Given that we assume only a small rotational broadening of the lines
for this star, a computation of three-component average profile was
found to be necessary. After a set of experiments, theoretical lines
from the mask were divided into three groups, according to their
predicted relative strengths: ``weak'' lines with individual
strengths $r$ (in units of continuum) between 0 and 0.4,
``intermediate'' strength lines with 0.4$<$$r$$\leq$0.6, and
``strong'' lines satisfying condition of $r$$>$0.6. In the next
step, the LSD profile is used to perform line strengths correction
for the mask that has been initially used for the calculation of the
mean (multicomponent) profile, and the procedure repeats until no
(significant) changes in the profile nor in the individual line
strengths occur. We then use the final list of the line strengths
and the LSD profile to compute a model spectrum and analyse it to
determine the fundamental parameters which are then compared to the
originally assumed ones. For estimation of the fundamental
parameters we rely on the spectral synthesis method as implemented
in the GSSP (Grid Search in Stellar Parameters) code
\citep{Tkachenko2012a,Lehmann2011}. The code finds the optimum
values of \te, \logg, $\xi$, $[M/H]$, and \vsini\ from the minimum
in $\chi^2$ obtained from a comparison of the ``observed'' spectrum
with the synthetic ones computed from all possible combinations of
the above mentioned parameters. The errors of measurement (1$\sigma$
confidence level) are calculated from the $\chi^2$ statistics using
the projections of the hypersurface of the $\chi^2$ from all grid
points of all parameters onto the parameter in question. For this
particular case, errors of measurement were estimated to be 75~K,
0.1~dex, 0.25~\kms, 0.7~\kms, and 0.08~dex in \te, \logg, $\xi$,
\vsini, and [M/H], accordingly.

\begin{figure}[t]
\includegraphics[scale=0.92]{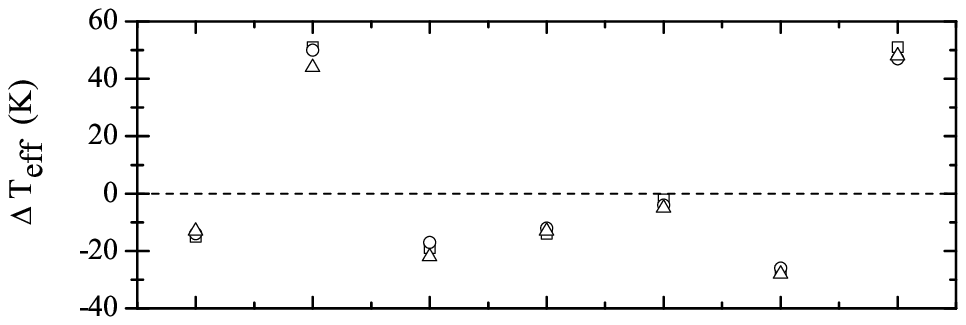}
\includegraphics[scale=0.92]{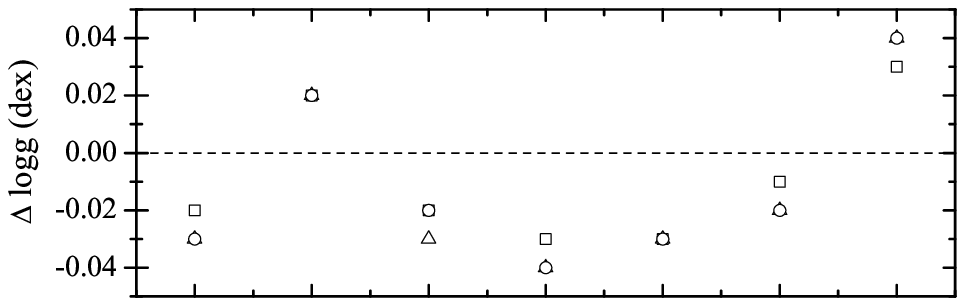}
\includegraphics[scale=0.92]{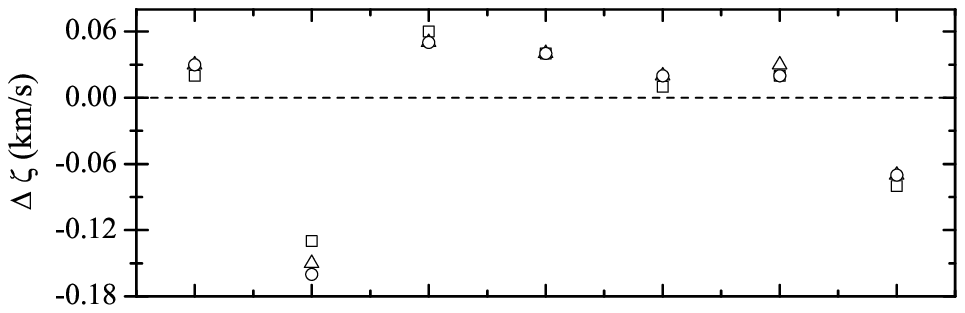}
\includegraphics[scale=0.92]{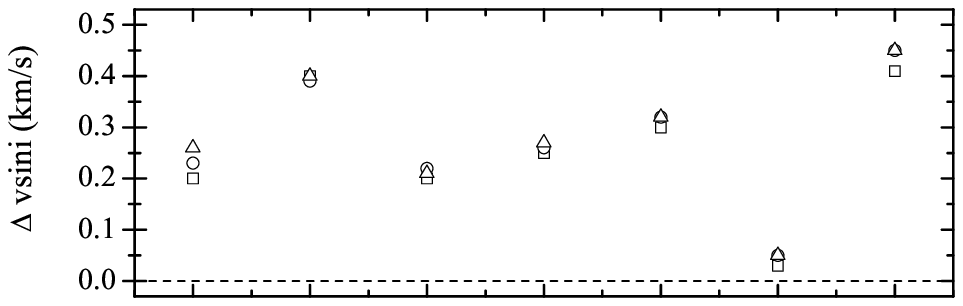}
\includegraphics[scale=0.92]{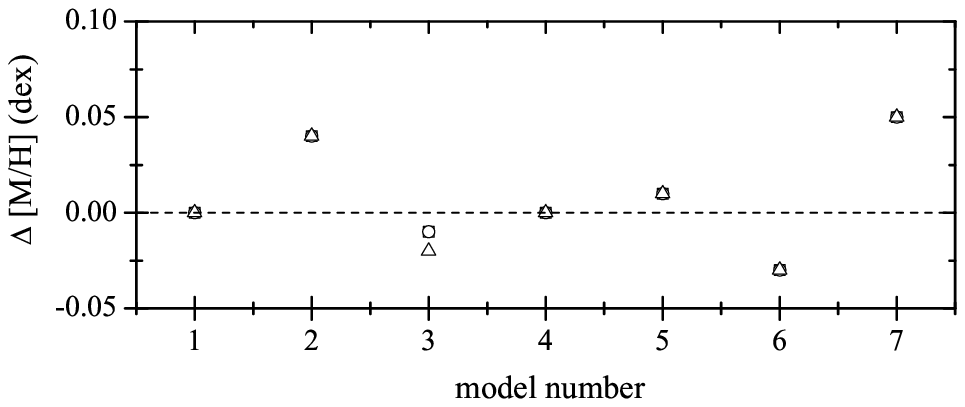}
\caption{{\small Comparison between the determined and the initial
model parameters for different S/N of the ``observed'' spectrum:
infinity (open squares), 70 (open circles), and 35 (open triangles).
Horizontal dashed line represents a perfect match. See text for
details.}} \label{Figure5}
\end{figure}

\begin{table}
\tabcolsep 1.9mm\caption{Model parameters used for the calculation
of different line masks. All models assume $\xi=$2~\kms. The
parameter that differs from the first model is given in
boldface.}\label{Table1}
\begin{tabular}{llcr|llcr} \hline\hline
model\rule{0pt}{9pt} & \te & \logg & [M/H] & model & \te & \logg & [M/H] \\
 & (K) & \multicolumn{2}{c|}{(dex)} & & (K) & \multicolumn{2}{c}{(dex)} \\\hline
1 & 8\,500\rule{0pt}{9pt} & 3.5 & 0.0 & 4 & 8\,500 & {\bf 3.0} & 0.0 \\
2 & {\bf 8\,000} & 3.5 & 0.0 & 5 & 8\,500 & {\bf 4.0} & 0.0 \\
3 & {\bf 9\,000} & 3.5 & 0.0 & 6 & 8\,500 & 3.5 & {\bf -0.3} \\
  &        &     &     & 7 & 8\,500 & 3.5 & {\bf +0.3} \\
\hline
\end{tabular}
\end{table}

\begin{figure}[t]
\includegraphics[scale=0.93]{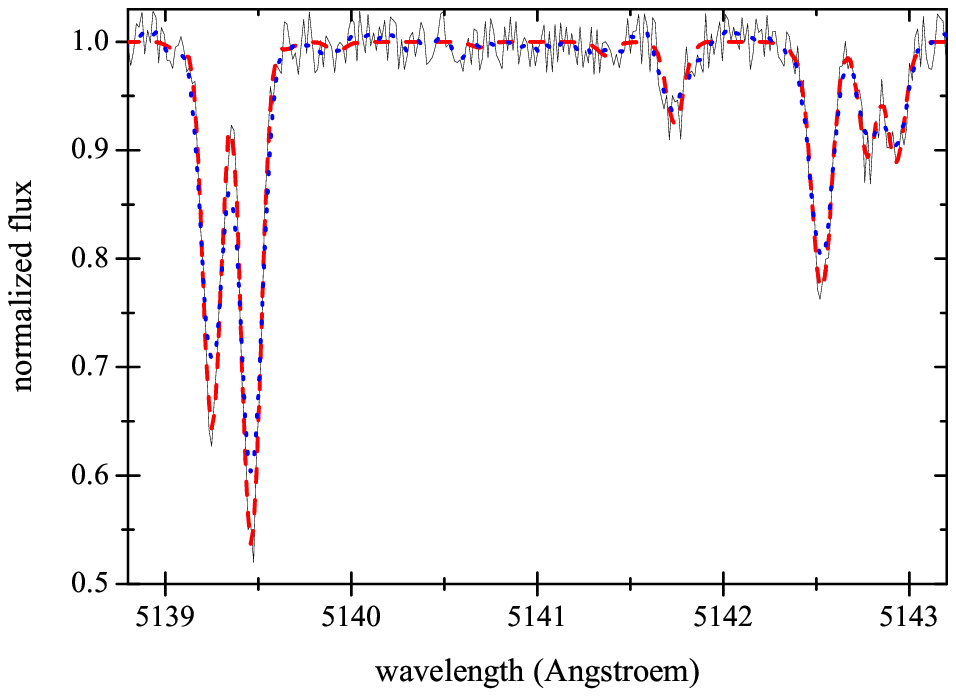}
\caption{{\small A few spectral features in the spectrum of a
synthetic single star (solid, black line). The smoothed version of
the spectrum (dotted, blue line) and the LSD model (dashed, red
line) are overplotted.}}\label{Figure6}
\end{figure}

Atmospheric parameters used for the calculation of the line masks
range between 8\,000--9\,000~K with step of 500~K in \te,
3.0--4.0~dex with step of 0.5~dex in \logg, and -0.3--+0.3~dex with
step of 0.3~dex in [M/H]. Figure~\ref{Figure5} summarizes the
results in terms of a difference between the values obtained from
spectral analysis of the final model spectrum and the initial
parameters. The x-axis gives the model sequence number,
corresponding parameters can be found in Table~\ref{Table1}. All
fundamental parameters are well reconstructed from our model spectra
and individual differences do not exceed 60~K, 0.04~dex, 0.18~\kms,
0.5~\kms, and 0.05~dex in \te, \logg, \vsini, $\xi$, and [M/H],
respectively. Such discrepancy appears to be perfectly within the
estimated errors of measurement. There is also no principle
difference between the spectra of different S/N values which is not
of a big surprise given that there are a couple of thousands
metallic lines in the considered wavelength range.

To check whether the rotation has an impact on the method and hence
the final results, we repeated the whole procedure for the same type
of star but assuming faster rotation with \vsini\ of 30 and then
60~\kms. The obtained results are the same as for the above
described slowly rotating star: all reconstructed atmospheric
parameters agree within the error bars with the initial values. This
leads us to the conclusion that the method is robust for both slowly
and rapidly rotating stars.

To verify whether the LSD-based analysis of stellar spectra is also
applicable to stars with individual abundances of some chemical
elements significantly deviating from the overall metallicity value,
we simulated spectra of a fake stellar object assuming
overabundances of Si and Mg of 0.5 and 0.2~dex, respectively. As in
the previous case, we tested the method for three different values
of S/N of the ``observed'' spectrum and checked how sensitive the
technique is to the initial model by using the same line masks as
for the normal star (see Table~\ref{Table1}). The obtained results
are essentially the same meaning that we can successfully
reconstruct back all the fundamental parameters and individual
abundances of the two chemical elements. An exception is the value
of an overall metallicity which seems to be largely affected by the
initially assumed overabundances of the two chemical elements and is
found to be by 0.07--0.09~dex higher than the initial parameter
value.

Finally, we compared our technique with a couple of widely used
smoothing algorithms, like running average and median filter. These
techniques are usually used to smooth out short-term fluctuations
(noise in case of spectra) and highlight long-term ``trends'' (read,
spectral features). Both techniques are equivalents of lowering
spectral resolution as they assume a replacement of several
neighbouring points by a single value, either average or median one.
Such smoothing is justified for the spectra of constant stars with
large values of projected rotational velocity. In such cases,
spectral features appear to be broad and lowering resolution has no
or a little influence on the derived characteristics of a star. The
situation is totaly different in the case of stars exhibiting narrow
lines in their spectra or/and intrinsically variable stars. In both
cases, smoothing is expected to have large impact on spectral
characteristics of a star, which is by artificial broadening of
spectral lines in the former case and by smoothing out the pulsation
signal in the latter. Indeed, Figure~\ref{Figure6} illustrating
simulated data of a slowly rotating star, clearly shows how the
spectral features can be artificially broadened by smoothing
algorithms. Estimated value of projected rotational velocity based
on these smoothed spectra is found to be a factor of 2.5 larger than
the original one. This in turn leads to incorrect values of other
fundamental parameters, which is not the case when using our
improved LSD algorithm.

\subsection{The cases of Vega and KIC04749989}

After testing the method on simulated data, we apply it to the
observed spectra of Vega and KIC04749989, with the aim to check
whether the obtained fundamental parameters agree with those
reported in the literature. The data were obtained with the {\sc
HERMES} spectrograph \citep{Raskin2011} attached to the 1.2-m
Mercator telescope (La Palma, Canary Islands). The spectra have a
resolving power of 85\,000 and cover a wavelength range from 380 to
900 nm. The data have been reduced using the dedicated pipeline,
including bias and stray-light subtraction, cosmic rays filtering,
flat fielding, wavelength calibration by ThArNe lamp, and order
merging. The continuum normalization was done afterwards by fitting
a (cubic) spline function through some tens of carefully selected
continuum points. More information on the adopted normalization
procedure can be found in \citet{Papics2012}.

\begin{figure}[t]
\includegraphics[scale=0.85]{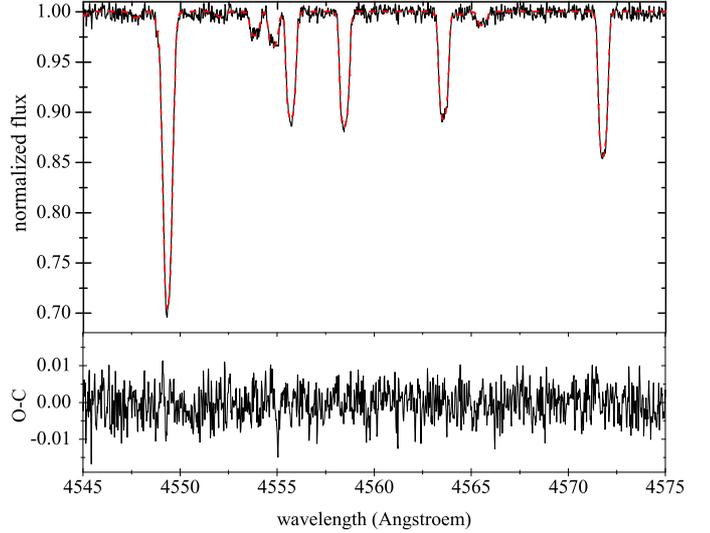}
\caption{{\small Comparison between the observed (solid, black line)
and final LSD model (dashed, red line) spectrum of Vega.}}
\label{Figure7}
\end{figure}

The LSD profiles for Vega were computed from several metal line
regions spread over the entire wavelength range of the {\sc HERMES}
spectra, and by excluding all hydrogen and helium spectral lines.
Figure~\ref{Figure7} compares two small parts of the observed
spectrum of the star with the final LSD-based model obtained after
the application of our line strengths correction algorithm. Similar
to the results obtained for the simulated data, the model matches
the observations very well and the obtained S/N is a factor of 5
higher than in the original spectrum. We used the obtained model
spectrum (free of Balmer and helium lines) to estimate the
fundamental parameters of the star by means of our GSSP code.
Table~\ref{Table2} compares the atmospheric parameters of Vega
obtained by us with those found by \citet{Lehmann2011} using the
same atmosphere models and spectrum synthesis codes as we did.
Obviously, all five atmospheric parameters are in perfect agreement
with the values reported by \citet{Lehmann2011}, which in turn agree
very well with several recent findings for this star \citep[see,
e.g., Table 3 in][]{Lehmann2011}. It is worth noting that the errors
on both \te\ and \logg\ are slightly higher than those estimated by
\citet{Lehmann2011}. This result also seems to be reasonable as in
our analysis we rely purely on metal lines, without using Balmer
profiles which are a good indicator of the effective temperature and
surface gravity in this temperature range.

Since we position our method as a good tool of increasing S/N of
stellar spectra without losing and/or affecting relevant
information, bright stars are not sufficient to test the technique
on: the original observed spectrum of Vega is of S/N$\sim$300 and
there is no need to improve the quality of the data to perform a
detailed analysis of the star. We thus tested the method on
KIC04749989, which contrary to Vega has a high value of \vsini\ of
$\sim$190~\kms\ and for which a {\sc HERMES} spectrum of S/N$\sim$60
has been obtained. The high value of the projected rotational
velocity for this star implies that all metal lines in its spectrum
are rather weak and shallow, and show very high degree of blending.

\begin{table} \tabcolsep 1.9mm\caption{Comparison of the atmospheric
parameters of Vega obtained by us with those reported by
\citet{Lehmann2011}. Errors of measurement estimated from $\chi^2$
statistics and represented by 1-$\sigma$ level are given in
parenthesis in terms of last digits.}\label{Table2}
\begin{tabular}{llll} \hline\hline
Parameter\rule{0pt}{9pt} & Unit & This paper &
\citet{Lehmann2011}\\\hline \te\rule{0pt}{9pt} & K & 9\,500 (140) &
9\,540 (120)\\
\logg & dex & 3.94 (12) & 3.92 (7)\\
\vsini & \kms & 22.3 (1.2) & 21.9 (1.1)\\
$\xi$ & \kms & 2.20 (40) & 2.41 (35)\\
${\rm [M/H]}$ & dex & --0.60 (12) & --0.58 (10)\\\hline
\end{tabular}
\end{table}

\begin{figure}[t]
\includegraphics[scale=0.85]{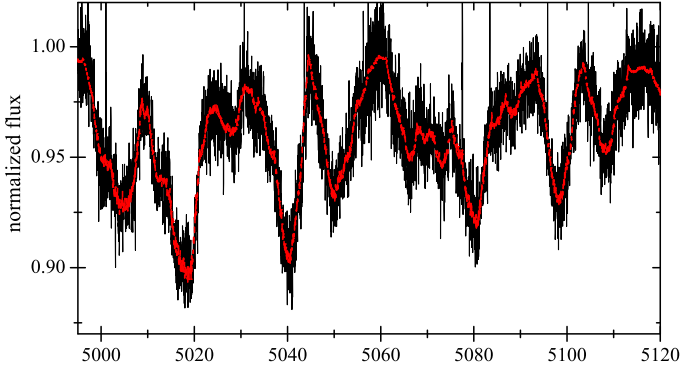}
\caption{{\small Same as Fig.~\ref{Figure7} but for KIC04749989.}}
\label{Figure8}
\end{figure}

For the calculation of the LSD profile and the corresponding model
spectrum, we used the wavelength region rich of metal lines, from
495 to 580 nm. Figure~\ref{Figure8} illustrates a part of the
observed spectrum of KIC04749989 (solid, black line) and compares it
with our final LSD-based model spectrum (dashed, red line). The
latter has S/N of $\sim$700 implying a gain by a factor of 10
compared to the original data. To check the impact of the
observational noise on the fundamental parameters estimate, both the
original and the LSD-based model spectra were analysed with the GSSP
code in the same, above mentioned wavelength region.
Table~\ref{Table3} compares the results revealing rather large
discrepancies between the two sets of obtained parameters. Indeed,
there is a difference of about 400~K and 0.3~dex in the effective
temperature and surface gravity, respectively, as well as
non-significant difference of 6~\kms\ in \vsini. As expected, the
rather high noise level in the original data has large impact on the
derived fundamental parameters which is also confirmed by the large
error bars. On the other hand, at least one properly normalized
Balmer profile significantly helps in constraining both \te\ and
\logg, which in turn leads to a more precise estimation of the two
other parameters. Indeed, the same observed, low S/N spectrum of
KIC04747989 was analysed by \citet{Tkachenko2013}, but using a wider
wavelength range from 470 to 580 nm, that, besides metal lines, also
covers the H$_{\beta}$ profile. The corresponding fundamental
parameters are listed in the last column of Table~\ref{Table3} and
agree very well with those derived by us from the high S/N LSD-based
model spectrum. Normalization of Balmer lines in noisy spectra is
far from trivial and is an additional source of uncertainties in the
derived \te\ and \logg. Our method does not rely on Balmer lines but
is fully based on fitting the metal lines wavelength regions, and
thanks to the significantly improved S/N of spectra, delivers
appropriate results.

Though there is no obvious advantage in using our proposed LSD-based
method for stars like Vega for which good quality data could be
obtained even with very small telescopes, the method promises to be
very robust for faint stellar objects (primary targets of, e.g., the
$Kepler$ spacecraft) observed with 1 to 2-m class telescopes. Given
that our LSD model spectra contain all relevant information present
in the original, observed spectra but provides much better S/N data,
the technique should also have its practical application to
intrinsically variable (e.g., rotationally modulated and/or
pulsating) stars. Thus, in the next Section, we investigate the
robustness of our method for pulsating stars, by testing it on
high-resolution spectra of 20~CVn and HD\,189631, two well-studied
\DSct\ and \GD-type pulsators.

\subsection{Asteroseismic targets: the cases of 20~CVn and HD\,189631}

As long as a star shows pulsations, asteroseismology, the study of
stellar interiors via interpretation of pulsation patterns observed
at the surfaces of the stars, allows for extremely accurate
estimation of fundamental stellar parameters like mass, radius,
density, etc. Knowledge of these parameters is in turn of major
importance for understanding stellar evolution as well as for the
correct characterization of exoplanets and their interaction with
host stars. Nowadays, in the era of the MOST \citep[Microvariability
and Oscillations of STars,][]{Walker2003}, CoRoT \citep[Convection
Rotation and Planetary Transits,][]{Auvergne2009} and $Kepler$
\citep{Gilliland2010} space missions providing us with
micro-magnitude precision photometric data for numerous pulsating
stars, we still lack additional information to perform in-depth
asteroseismic studies for many of those stars. Indeed, the above
mentioned missions observe their targets in white light which makes
identification of individual pulsation modes, one of the
pre-conditions for asteroseismology to work, impossible from these
data unless one can identify the modes from frequency or period
patterns alone. This is usually the case for stochastically excited
modes in the asymptotic acoustic regime or for high-order gravity
mode pulsations, but not in general for heat-driven modes
\citep[e.g.,][]{Aerts2010}. For the latter, one needs extra
information in terms of time-series of (ground-based) multicolour
photometry and/or high-resolution spectroscopy to be obtained. A
combination of the high-resolution but moderate S/N spectra obtained
with lower class telescopes and our new technique, may provide data
of sufficient quality to perform the desired analysis.

In this section, we test our method on 20 Canum Venaticorum and
HD\,189631, a bright ($V$=4.73) \DSct-type pulsator and a
non-radially pulsating \GD\ star, respectively. Our ultimate goal is
to check whether the LSD-based model spectra providing a much higher
S/N compared to the original data, at the same time, do not affect
the oscillation signal present in the original spectra in terms of
LPV. This is done by extracting individual frequencies from the
time-series of the LSD model spectra and performing identification
of the individual oscillation modes. The results are then compared
to those obtained from the original data as well as to the results
of previous studies. In both cases, we make use of the {\sc FAMIAS}
software package \citep{Zima2008} for the extraction of the
individual frequencies as well as for the mode identification. The
analysis is based both on the moment \citep{Aerts1992,Briquet2003}
and pixel-by-pixel \citep{Zima2008} methods.

\begin{table}[t]
\tabcolsep 1.9mm\caption{Comparison of the atmospheric parameters of
KIC04749989 obtained by us from the original data and from the
LSD-based model spectrum with those reported by \citet[][indicated
as T2013]{Tkachenko2013}. The latter study made use of the
H$_{\beta}$ profile to put additional constraints on \te\ and \logg,
and is used here as a reference. Errors of measurement estimated
from $\chi^2$ statistics and represented by 1-$\sigma$ level are
given in parenthesis in terms of last digits.}\label{Table3}
\begin{tabular}{lllll} \hline\hline
Parameter\rule{0pt}{9pt}&Unit & \multicolumn{2}{c}{This paper} & \multicolumn{1}{c}{T2013}\\
& & \multicolumn{1}{c}{Original} & \multicolumn{1}{c}{LSD model}
&\\\hline \te\rule{0pt}{9pt} & K & 6\,950 (200) & 7\,380 (130) & 7\,320 (120)\\
\logg & dex & 4.02 (45) & 4.34 (25) & 4.32 (35)\\
\vsini & \kms & 184 (10) & 190 (8) & 191 (10)\\
$\xi$ & \kms & 2$^*$ & 2$^*$ & 2$^*$\\
${\rm [M/H]}$ & dex & +0.12 (12) & --0.03 (8) & +0.00 (12)\\\hline
\multicolumn{5}{l}{$^*$ fixed\rule{0pt}{9pt}}
\end{tabular}
\end{table}

\subsubsection{20 Canum Venaticorum}

Periodicity of 20 CVn was discovered by \citet{Merrill1922} who
reports it to be a binary candidate. However, \citet{Wehlau1966}
noticed this misclassification of the star as a binary and propose
the \DSct-type oscillations as the cause of the observed
variability. Since then, 20 CVn was the subject of several
photometric and spectroscopic studies which led to a large diversity
in results and interpretations. For example, \citet{Shaw1976} finds
the star to be a mono-periodic \DSct\ pulsator whereas
\citet{Smith1982} and \citet{Bossi1983} report the detection of a
second oscillation mode. The same holds for the proposed type of
oscillations where, for instance, \citet{Mathias1996} classify the
star as a non-radial pulsator whereas \citet{Rodriguez1998} report
on the radial identification for the observed pulsation mode. The
conclusions by \citet{Rodriguez1998} were later on confirmed by
\citet{Chadid2001} and \citet{Daszynska2003}. The atmospheric
parameters are also not very well constrained for this star
revealing a scatter in the literature of about 700~K, 0.7~dex, and
10~\kms\ in \te, \logg, and \vsini, respectively \citep[see,
e.g.,][]{Hauck1985,Rodriguez1998,Erspamer2003}. This particular
diversity in the reported fundamental parameters provides an
excellent test of our technique and, in particular, allows to check
an impact of the selected line mask on the final results.

We base our analysis on spectra taken with the {\sc ELODIE}
\'echelle spectrograph, at the time of its operation (1993--2006)
attached to the 193-cm telescope at Observatoire de Haute-Provence
(OHP). The data have a resolving power of 42\,000, cover wavelength
range from 3895 to 6815~\AA, and were originally obtained by
\citet{Mathias1996}, \citet{Chadid2001}, and \citet{Erspamer2003}.
We used the {\sc ELODIE} archive \citep{Moultaka2004} to retrieve
the data, Table~\ref{Table4} gives the journal of observations
listing the source, the period of observation, number of the
acquired spectra, and the average S/N value.

The {\sc ELODIE} archive contains pre-extracted, wavelength
calibrated spectra. Besides the continuum normalization and
correction for the Earth's motion, we carefully examined the spectra
for any additional systematic nightly RV variations. The variability
of the order of 100-300~m\,s$^{-1}$ was detected, in perfect
agreement with the findings by \citet{Chadid2001}. These authors
attribute this variability to instabilities of the {\sc ELODIE}
spectrograph, more specifically, of the ThAr spectra served to
calibrate the spectral wavelength scaling. Following the procedure
proposed by \citet{Chadid2001}, we corrected the spectra from each
night by matching the corresponding (nightly) mean RV to the mean
value observed in 1998.

\begin{figure}
\includegraphics[scale=0.37]{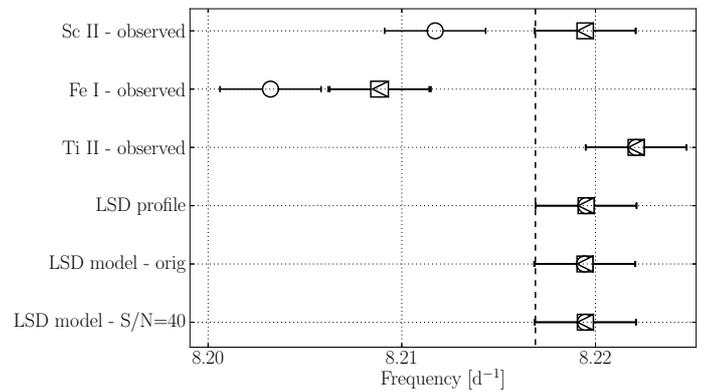}
\caption{{\small Comparison of the frequencies extracted from the
three different spectral lines in the original observed spectra of
20\,CVn with the values deduced both from the LSD-based model
spectra (for the Sc\,{\small II}\,5239~\AA\ only, based on the data
sets characterized by different S/N) and the LSD profiles
themselves. The symbols refer to the three observables used for the
analysis: the first- and the third-order moment (squares and
circles, respectively), and the profiles themselves (triangles). The
frequency errors are assumed to correspond to the Rayleigh limit of
the data. The vertical dashed line represents the literature value
of 8.2168~\cd\ \citep{Rodriguez1998}.}} \label{Figure9}
\end{figure}

\begin{table}
\tabcolsep 2.5mm\caption{\label{Table4} Journal of observations for
20~CVn and HD\,189631. Besides the original source and the period of
observations, we list the number of obtained spectra ($N$) and the
average value of S/N.}
\begin{tabular}{llrr} \hline\hline\
Source\rule{0pt}{9pt} & \multicolumn{1}{c}{Observing period} & $N$ & S/N\\
\hline \multicolumn{4}{c}{{\bf 20~CVn}\rule{0pt}{9pt}}\\
\citet{Mathias1996}\rule{0pt}{9pt} & 20--21.12.1994 & 8 & 84\\
\citet{Chadid2001} & 11--15.04.1998 & 113 & 127\\
 & 24--25.04.1999 & 31 & 168\\
 & 27--30.04.1999 & 131 & 126\\
\citet{Erspamer2003} & 05--06.06.1999 & 1 & 308\\
Unknown & 28--29.03.2002 & 1 & 260\\
\multicolumn{4}{c}{{\bf HD\,189631}\rule{0pt}{9pt}}\\
\citet{Maisonneuve2011}\rule{0pt}{9pt} & 02--09.07.2008 & 274 & 215\\
& 22.06--02.07.2009 & 24 & 165\\
& 17--21.07.2009 & 25 & 360\\\hline
\end{tabular}
\end{table}

\begin{figure*}
\centering
\includegraphics[scale=0.45]{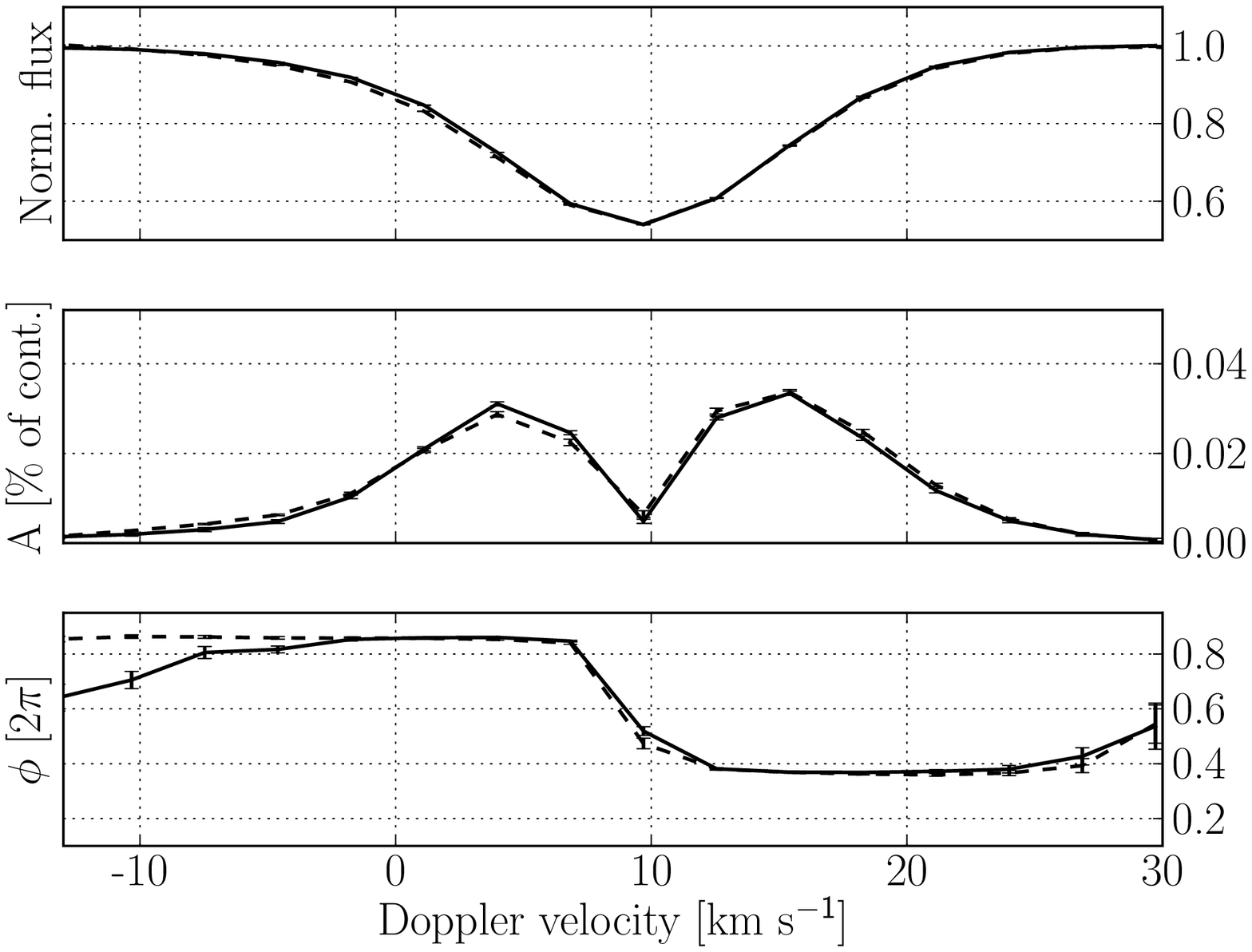}\hspace{3mm}
\includegraphics[scale=0.45]{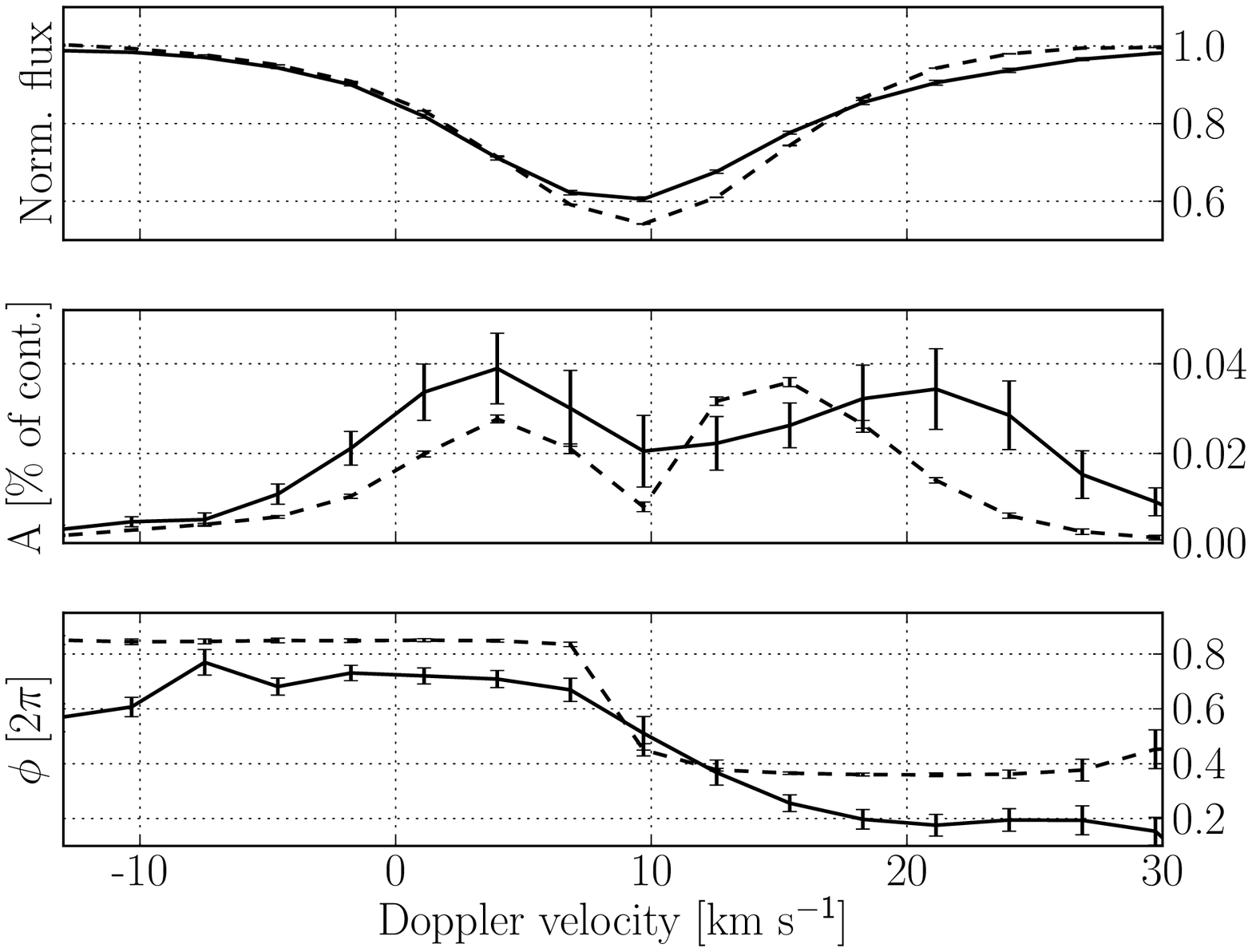}
\caption{{\small Results of the frequency analysis of Sc\,{\small
II}\,5239~\AA\ line in the spectra of 20\,CVn by means of the
pixel-by-pixel method. The two panels stand for the original data
(left) and the simulated data set characterized by S/N of 40
(right). Each panel illustrates the mean profile (top), and the
amplitude and phase across this profile (middle and bottom,
respectively). The solid and dashed lines represent the original
observed and the LSD-based model spectra, respectively. The
corresponding error margins are shown by the bars. The results are
shown for the dominant (radial) mode of $\sim$8.22~\cd\
(95.11~\mhz).}} \label{Figure10}
\end{figure*}

As can be seen from the last column of Table~\ref{Table4}, the
original data are of rather good quality and have an average S/N
above 100. In order to test the influence of the noise on the final
results, we created an additional data set by adding noise to the
original spectra of 20 CVn to simulate S/N of 40. Both data sets
were subject for frequency analysis which was performed based on
three rather strong and weakly blended spectral lines: Sc\,{\small
II}\,5239~\AA, Fe\,{\small I}\,5288~\AA, and Ti\,{\small
II}\,5336~\AA. The obtained results were then used for comparison
with both the literature and the results obtained from the LSD-based
model spectra. The latter were computed from both data sets
(original and those with S/N of 40), assuming a line mask
characterized by \te=7\,600~K, \logg=3.6~dex, and [M/H]=+0.5~dex,
the parameters reported by \citet{Rodriguez1998}. We based our
analyses on the pixel-by-pixel method and on the first and the third
moments of the line profiles, whereas we were unable to detect any
significant variability in their second moment.

\begin{table*}[ht]
\centering \tabcolsep 3.0mm\caption{\label{Table5} Results of the
frequency analysis for HD\,189631. The values obtained by
\citet{Maisonneuve2011} are given for comparison. The Rayleigh limit
amounts to 10$^{-3}$~\cd\ and 10$^{-5}$~\cd\ for the data used in
this paper and those of \citet{Maisonneuve2011}, respectively.
$\langle v\rangle$, $\langle v^3\rangle$, and PBP stand for the
first and the third moments, and the pixel-by-pixel method,
respectively.}
\begin{tabular}{llllllllll} \hline\hline
\multirow{2}{*}{Freq}. & \multicolumn{1}{c}{\multirow{2}{*}{Unit}} & Maisonneuve et al.\rule{0pt}{9pt} & \multicolumn{2}{c}{Identification} & \multicolumn{3}{c}{This paper} & \multicolumn{2}{c}{Identification}\\
& & \multicolumn{1}{c}{(2011)} & \multicolumn{1}{c}{$l$} & \multicolumn{1}{c}{$m$} & \multicolumn{1}{c}{$\langle v \rangle$} & \multicolumn{1}{c}{$\langle v^3 \rangle$} & \multicolumn{1}{c}{PBP} & \multicolumn{1}{c}{$l$} & \multicolumn{1}{c}{$m$}\\
\hline
f$_1$\rule{0pt}{9pt} & \multirow{4}{*}{\cd\ (\mhz)} & 1.6719 (19.3439) & \multicolumn{1}{c}{1} & \multicolumn{1}{c}{1} &1.682 (19.461) & 1.682 (19.461) & 1.685 (19.495) & \multicolumn{1}{c}{1} & \multicolumn{1}{c}{1}\\
f$_2$ & & 1.4200 (16.4294) & \multicolumn{1}{c}{1} & \multicolumn{1}{c}{1} & 1.382 (15.990) & 1.382 (15.990) & 1.411 (16.325) & \multicolumn{1}{c}{1} & \multicolumn{1}{c}{1}\\
f$_3$ & & 0.0711 (0.8226) & \multicolumn{1}{c}{2} & \multicolumn{1}{c}{--2} & 0.119 (1.377) & 0.127 (1.469) & 0.122 (1.412) & \multicolumn{1}{c}{---} & \multicolumn{1}{c}{---}\\
\multirow{3}{*}{f$_4$} & & \multirow{3}{*}{1.8227 (21.0886)} &
\multicolumn{1}{c}{4} & \multicolumn{1}{c}{1} &
\multirow{3}{*}{1.823 (21.092)} & \multirow{3}{*}{1.820 (21.057)} &
\multirow{3}{*}{1.826 (21.127)} & \multicolumn{1}{c}{\multirow{3}{*}{1}} & \multicolumn{1}{c}{\multirow{3}{*}{1}}\\
& & & \multicolumn{2}{c}{or} & & & & &\\
& & & \multicolumn{1}{c}{2} & \multicolumn{1}{c}{--2} & & &
&\\\hline
\end{tabular}
\end{table*}

The results obtained from the LSD-based model spectra were
essentially the same for all the three above mentioned spectral
lines, as they were the same from the analysis of average profiles
themselves. For this reason, from this point on, we focus on the
results obtained for the Sc\,{\small II}\,5239~\AA\ spectral line
only but the conclusions are valid for the two other lines and for
the average profiles. Figure~\ref{Figure9} compares the frequencies
obtained from the analysis of all the three lines in the original
observed spectra with those deduced from the LSD profiles themselves
as well as from the LSD-based model of the Sc\,{\small
II}\,5239~\AA\ line. As expected, the frequency obtained from the
analysis of this line in the LSD model spectra matches well the one
deduced from the average profiles themselves. This is because all
the information about LPV that is present in the LSD-model spectra
entirely comes from the average profiles extracted directly from the
observations. Our decision to work here with several individual
lines from the LSD spectra but not with the mean profiles themselves
is due to interest to what extend the noise can influence results of
the frequency analysis when applied to different observables. In
Figure~\ref{Figure9}, one can see that frequencies extracted from
the LSD model spectra (whatever the S/N value is assumed) using
different diagnostics (pixel-by-pixel or spectral line moments) are
in perfect agreement with each other and with the value reported by
\citet{Rodriguez1998} (vertical dashed line). On the other hand,
being applied to the original spectra, frequency analysis of the
third-order moment of the line (open circle in the top most line in
Figure~\ref{Figure9}) reveals different results from those obtained
by pixel-by-pixel and the first-order moment methods. Since the
degree of blending of this particular spectral line of Sc\,{\small
II} at 5239~\AA\ is the same both in the original and LSD model
spectra, we conclude that the above mentioned difference in the
extracted frequencies is purely due to noise present in the original
data. We also find that the Fe\,{\small I}\,5288~\AA, and
Ti\,{\small II}\,5336~\AA\ lines from the original observed spectra
deliver slightly different results, but this can be explained by the
higher degree of blending and effects of normalization to the local
continuum for these two lines. Finally, we stress that both the
moment and pixel-by-pixel methods applied to the LSD model spectra
or the average profiles themselves lead to consistent results for
this star (see different symbols in Figure~\ref{Figure9}).

Figure~\ref{Figure10} illustrates the results of the frequency
analysis of the Sc\,{\small II}\,5239~\AA\ line by means of the
pixel-by-pixel method, based on both the observed and the LSD-based
model spectra and assuming two different data sets: original (left)
and simulated, characterized by S/N of 40 (right). Each panel
contains three subplots showing the mean profile, and the amplitude
and phase variations corresponding to the determined frequency of
8.2195~\cd\ (95.0996~\mhz). The solid and dashed lines refer to the
observed and LSD model data, respectively, the corresponding error
margins are indicated by the bars. One can see that the resulting
mean profile as well as the amplitude and phase measured across this
profile are all strongly affected by the noise present in the
observations (solid lines and associated with them errors in the
right panel). This implies that the obtained observables based on
the spectra of S/N$=$40 cannot be used for further analysis and, in
particular, identification of individual oscillation modes. On the
other hand, the LSD-based model spectra computed from the same
observations provide us with much more accurate measurements which
appear to be in a good agreement with those obtained from the
original, high S/N data (left panel) and are thus well suitable for
further mode identification. Indeed, the identification of the
dominant oscillation mode of 8.2195~\cd\ (95.0996~\mhz) suggests it
to be either a radial mode or $l=1$ zonal mode. The latter
identification is favoured by the moment method providing slightly
lower reduced $\chi^2$ value, while the former identification as a
radial mode is found to be the most probable when relying on the
pixel-by-pixel method. This finding is in perfect agreement with the
previous studies by, e.g., \citet{Rodriguez1998} and
\citep[][Chapter 6]{Aerts2010}.

\begin{figure*}
\centering
\includegraphics[scale=0.45]{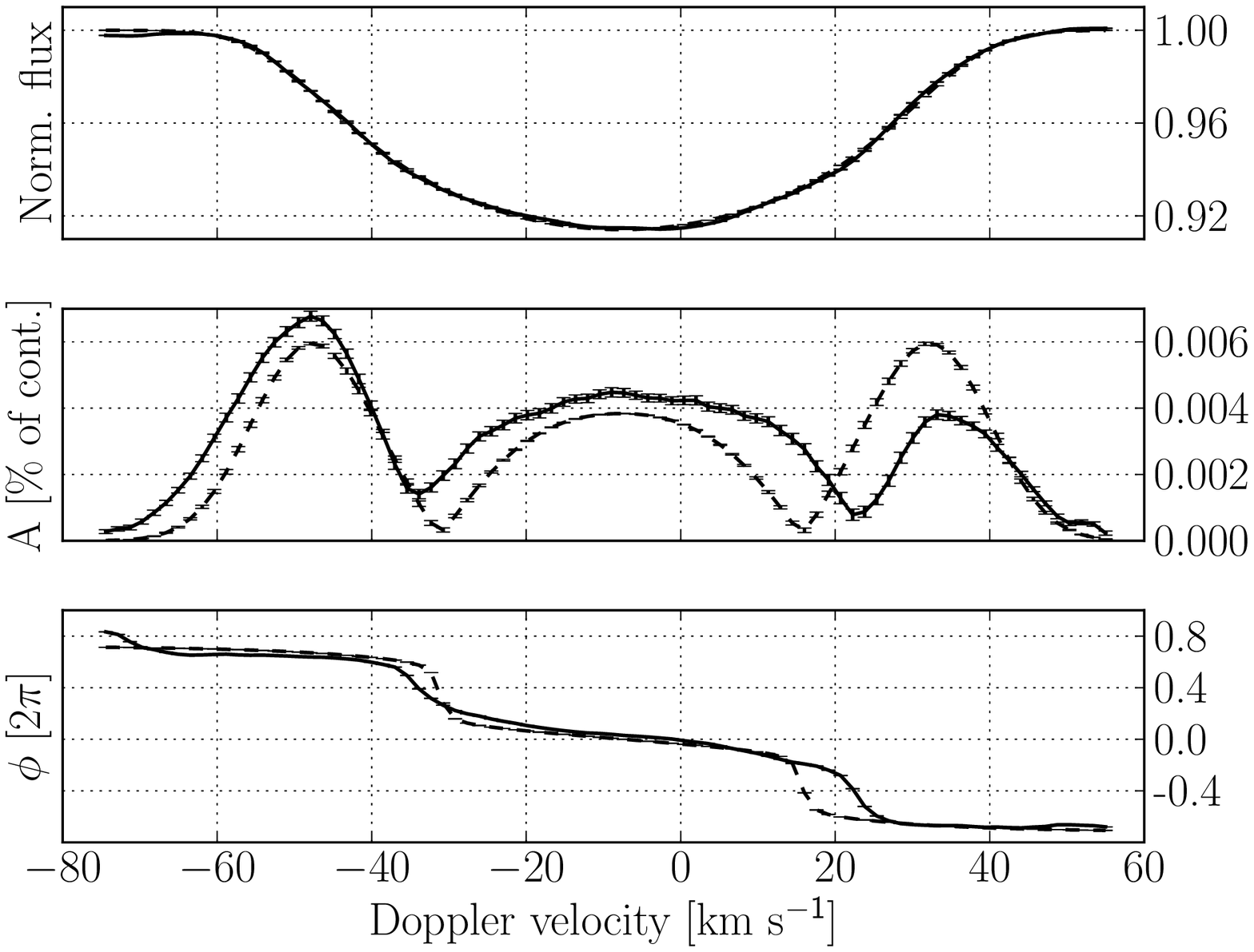}\hspace{3mm}
\includegraphics[scale=0.45]{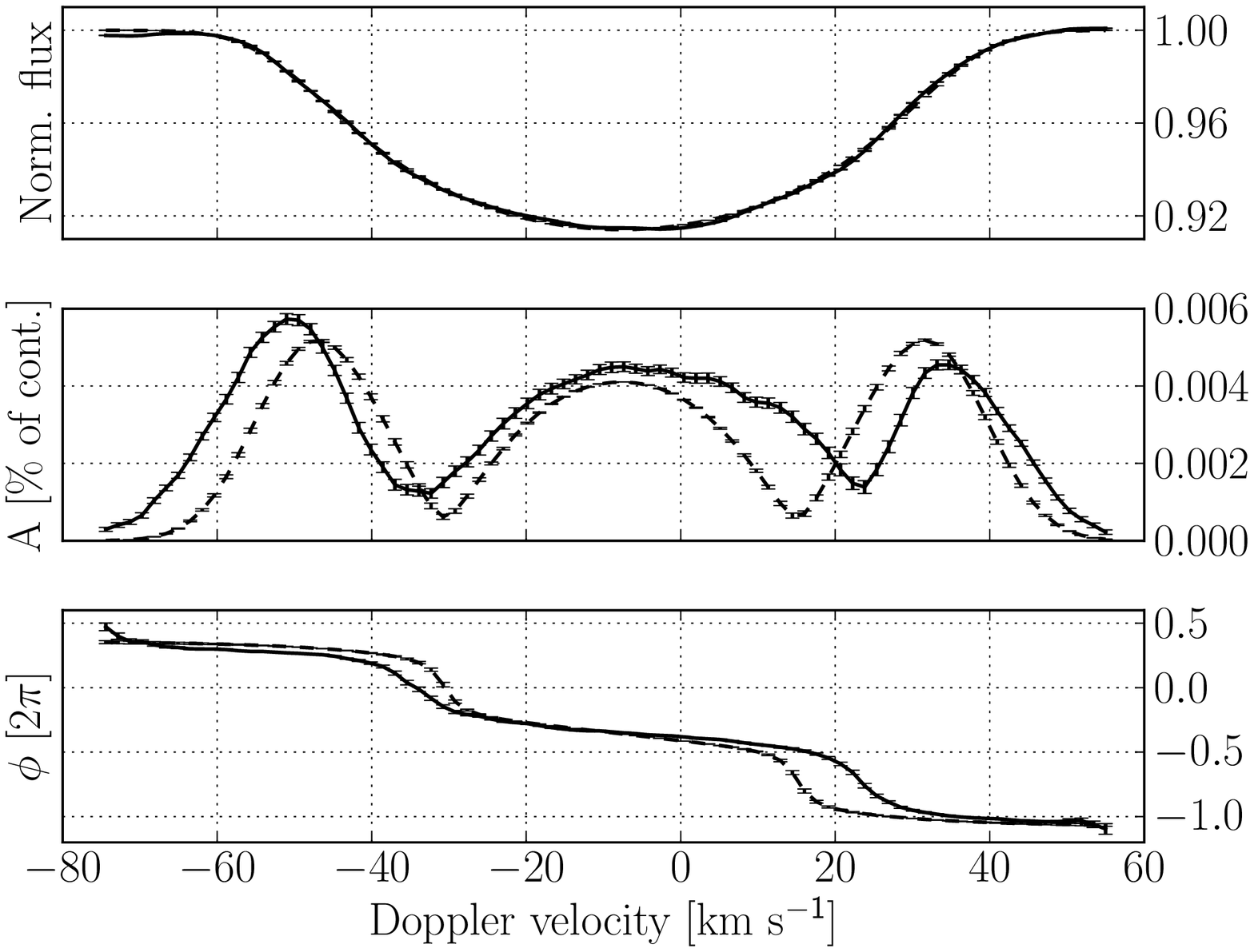}\vspace{5mm}
\includegraphics[scale=0.45]{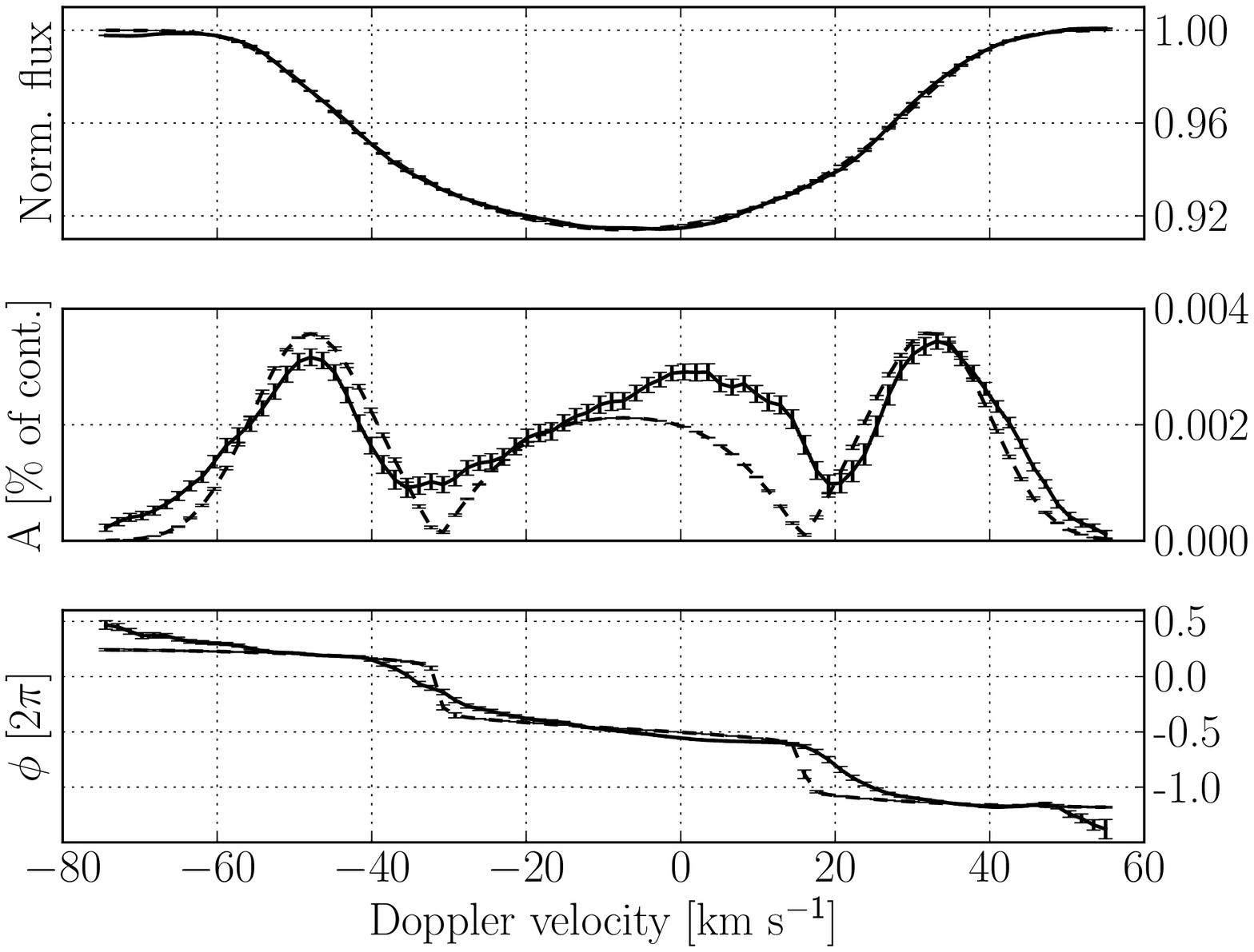}\hspace{3mm}
\includegraphics[scale=0.45]{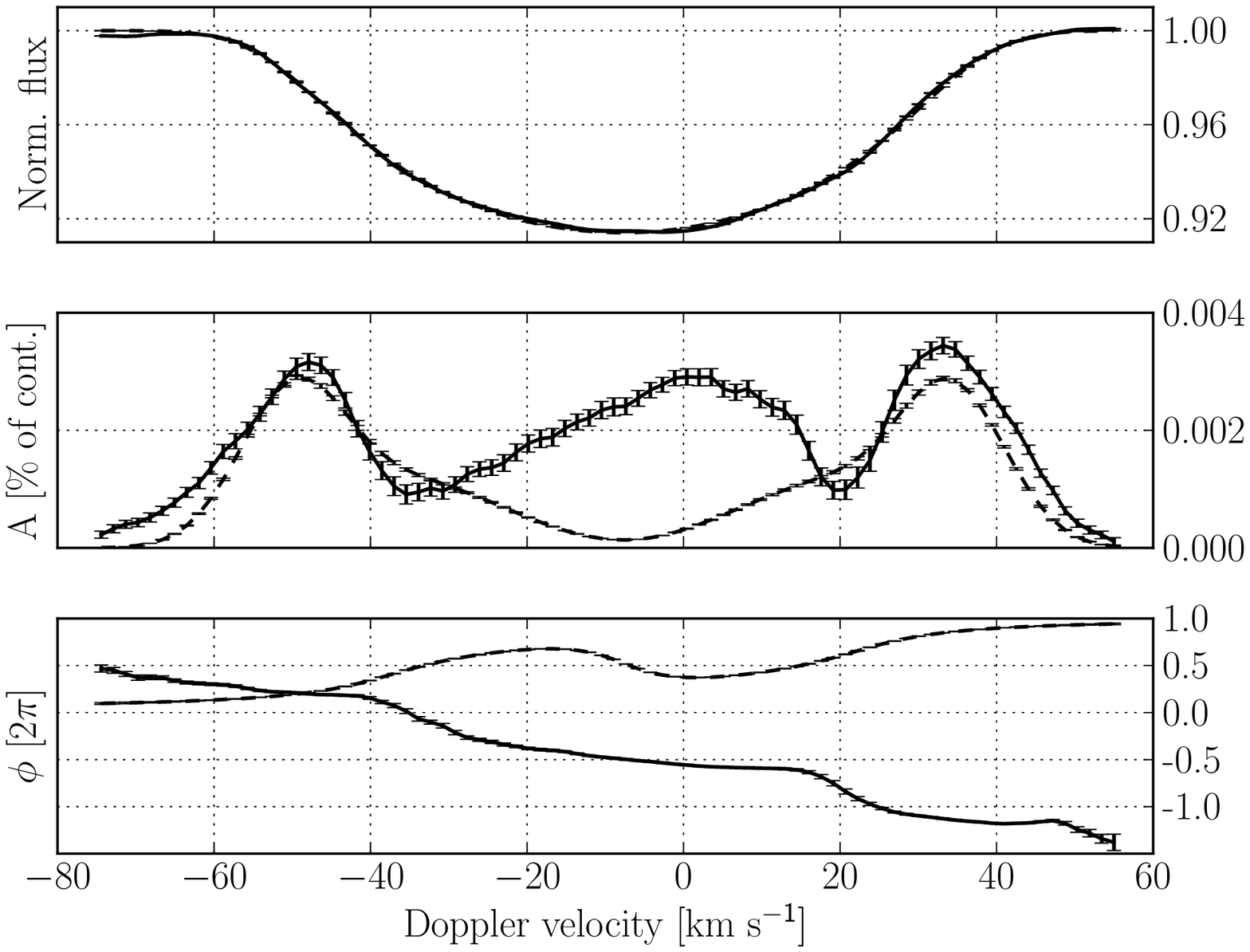}
\caption{{\small Results of the mode identification by means of the
pixel-by-pixel method for the three (out of four) frequencies
detected in the spectra of HD\,189631: f$_1$ (top left), f$_2$ (top
right), and f$_4$ (bottom). The bottom left and right panels stand
for the $(l,m)=(1,1)$ and $(l,m)=(2,-2)$ identifications,
respectively. The solid and dashed lines stand for the observations
(LSD spectra) and the ({\sc FAMIAS}) model, respectively. Both the
frequency values and their identifications are listed in
Table~\ref{Table5}.}} \label{Figure11}
\end{figure*}

\subsubsection{HD\,189631}

HD\,189631 is a bright ($V$=7.54), late A-type \GD\ non-radial
pulsator. We chose this object to test our LSD method on for the two
reasons: i) this star is one of a few \GD-type pulsators for which
an extensive spectroscopic mode identification study has been
conducted \citep{Maisonneuve2011}, and ii) more than 300
high-resolution spectra obtained by \citet{Maisonneuve2011} with the
{\sc HARPS} \'{e}chelle spectrograph attached to the ESO's 3.6-m
telescope (La Silla, Chile) were publicly available through the ESO
Science Archive. The obtained data were already pre-extracted and
wavelength calibrated, and additionally normalized to the continuum
by us. Table~\ref{Table4} gives the journal of observations listing
the observing period, the number of the obtained spectra, and the
average S/N.

Having an independent mode identification was important for us
because the star is a moderate rotator
\citep[$\sim$45~\kms,][]{Maisonneuve2011} showing a large number of
metal lines in its spectrum, which provides a high degree of
blending making mode identification based on single observed line
profiles a very difficult task. Thus, we used the original observed
spectra to compute the LSD profiles and analysed the latter to
determine frequencies and identification of the individual pulsation
modes. In the previous Section we showed that i) the LSD profiles
deliver essentially the same results as the LSD model spectra, and
ii) the noise level in the original spectra has no impact on the
final results and their interpretation given that sufficient number
of metallic lines are available for calculation of the average
profile. Thus, for this particular star, we restricted our analysis
to the LSD profiles only, computed for the original, high S/N ratio
spectra, and aimed to check whether the LSD-based analysis is also
suitable for multi-periodic, non-radially pulsating stars.

We compare our results to those reported by \citet{Maisonneuve2011}.
As such, it is worth noting that these authors benefitted from
additional spectroscopic data obtained with the {\sc HERCULES} and
{\sc FEROS} spectrographs operated at the Mount John University
Observatory (MJUO, New Zealand) and European Southern Observatory
(ESO, Chile), respectively. In practice, \citet{Maisonneuve2011} had
a longer time base of the observations which in turn transforms into
a higher frequency resolution. Similar to the case of 20~CVn, we
base our analysis on both the moment and the pixel-by-pixel methods.
The results of the frequency analysis are summarized in
Table~\ref{Table5}. Similar to \citet{Maisonneuve2011}, we detected
four individual frequency peaks, all in the typical \GD\ stars
low-frequency domain. Despite the fact that we had less data at our
disposal, the frequency values obtained by us by (at least) one of
the methods compare well to those reported by
\citet{Maisonneuve2011}, except for the f$_3$ mode. Thus, we focused
on the three modes f$_1$, f$_2$, and f$_4$, and performed a
simultaneous mode identification for them. We have found that all
the three modes are probably $l$=1 sectoral modes, the results for
f$_1$ and f$_2$ are consistent with the findings by
\citet{Maisonneuve2011}. The identification for f$_4$ is different
from the one presented by these authors, but it is worth noting that
the $(l,m)=(2,-2)$ geometry appears to be second best solution in
our calculations (the solution proposed by \citet{Maisonneuve2011}
as a possible one, see Table~\ref{Table5}).

Figure~\ref{Figure11} illustrates the results of the mode
identification for all the three frequencies by showing the fit to
the mean profiles, and the amplitude and phase variations. For the
highest frequency mode f$_4$, we present two identifications, of
which one corresponds to our best solution dipole mode (bottom
left), while the second one stands for $l=2$ retrograde sectoral
mode (bottom right) proposed by \citet{Maisonneuve2011} as one of
the possible solutions. Clearly, $l=2$ mode does not allow to fit
neither the observed amplitude nor phase variations across the
profile, with the largest discrepancy for the amplitude being in the
center of the line. The $(l,m)=(4,1)$ solution proposed by
\citet{Maisonneuve2011} for this mode, though not shown here, was
also found by us as one of the possible identifications. However, we
find it unlikely that f$_4$ is $(l,m)=(4,1)$ mode because (i)
according to the obtained $\chi^2$ values, this solution appears to
be twice as bad as the dipole mode, and (ii) the obtained
inclination angle is close to the so-called {\it inclination angle
of complete cancelation} \citep[IACC,
e.g.,][]{Chadid2001,Aerts2010}, which stands for the angle at which
a particular mode becomes undetectable in the integrated light due
to geometrical cancelation effect. Thus, we conclude that similar to
the two highest amplitude modes f$_1$ and f$_2$, the f$_4$ mode is
most probably a $l=1$ sectoral mode.

\begin{figure}[t]
\includegraphics[scale=0.87]{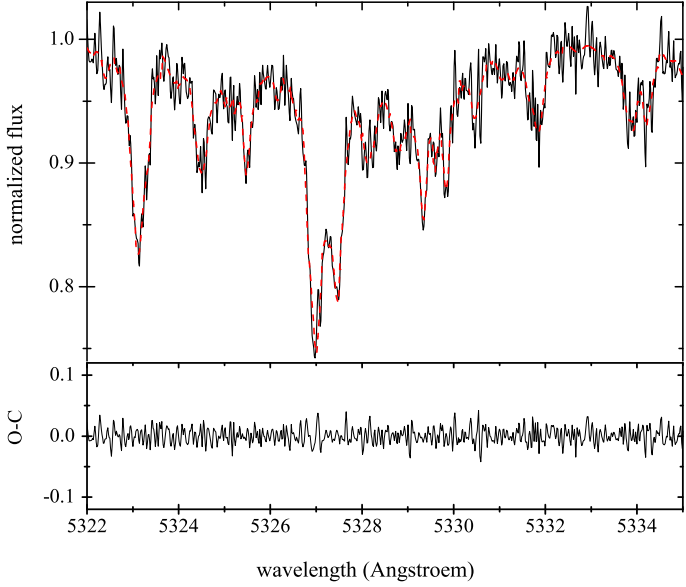}
\caption{{\small Same as Fig.~\ref{Figure7} but for
KIC\,11285625.}}\label{Figure12}
\end{figure}

The fact that all the three modes are found to have the same
spherical degree $l$ suggests that these might be members of either
rotationally splitted multiplet or (non)uniform period spacing of
gravity modes \citep[for detailed description of both effects, see
e.g.][]{Aerts2010}. Rotationally splitted modes tend to have the
same degree $l$ but different azimuthal number $m$ and the value of
splitting allows to deduce the rotational period of the star. Thus,
we checked whether at least one of the three considered here
oscillation modes could be characterized by negative value of $m$,
that is, being a retrograde mode. From the modelling, we find that
for all the three modes, negative sign for azimuthal number $m$
suggests either nearly zero amplitude in the center of the line or
increasing phase across the profile (similar to what is shown by the
dashed line in the bottom right panel of Figure~\ref{Figure11}).
This contradicts the observed characteristics of the modes, as all
of them show decreasing phase and exhibit significant amplitudes in
the line center (cf. Figure~\ref{Figure11}). Thus, we conclude that
the three modes have positive $m$ values and cannot be the members
of a rotational multiplet. On the other, the first-order asymptotic
approximation for non-radial pulsations predicts the periods of
high-order g-modes to be equally spaced
\citep{Tassoul1980,Aerts2010}. That is, low-degree gravity modes
characterized by the same $l$ and $m$ numbers but different radial
orders $n$ may form multiplets of equidistantly spaced in period
peaks. The spacings can also be non-equidistant if, e.g., the
chemical composition gradient exists near the stellar core
\citep[e.g.,][]{Miglio2008,Degroote2010}. The two of the detected
and identified in this work as $(l,m)=(1,1)$ modes of HD\,189631,
f$_1$ and f$_2$, show the period spacing which is $\sim$2.5 times
larger than the spacing between f$_1$ and f$_4$ modes. The latter
spacing of $\sim$3\,900~s seems to be well in the range of
theoretically predicted spacings of $l=1$ modes for a typical
$\gamma$\,Dor-type pulsator. Thus, we speculate that the three modes
f$_1$, f$_2$, and f$_4$ might show non-equidistant period spacing,
but also stress that an in-depth (theoretical) analysis is needed to
prove that. This kind of analysis is beyond the scope of the current
paper.

\section{Composite spectra of binary stars}\label{Section5}

As discussed in Section~3, the multiprofile technique, in particular
allows to account for difference in atmospheric conditions of the
two stars forming a binary system by assuming two different sets of
spectral lines in a mask. With a proper mask choice, similar to the
case of single stars, our improved LSD technique has a potential of
good representation of a binary composite spectrum and thus, its
effective denoising.

For spectroscopic double-lined binaries, determination of
fundamental parameters of both components is usually achieved by
means of the spectral disentangling technique
\citep{Simon1994,Hadrava1995}. The method requires a good orbital
phase coverage and delivers mean decomposed spectra that can be
analysed by means of the tools suitable for single stars. One also
benefits in S/N, the gain is proportional to $\sqrt N$ (additionally
scaled to the light ratio of the two stars), with $N$ the number of
individual composite spectra. With only a few rather noisy (but well
sampled in orbital phase) composite spectra, the gain in S/N is
usually not sufficient to conduct the desirable analysis. For
example, this was the case for KIC\,11285625, a spectroscopic
double-lined eclipsing binary with a \GD-type pulsating primary
component. Due to the originally noisy composite spectra and a small
contribution of the secondary component to the total light of the
system, the resulting decomposed spectrum of the secondary was of
insufficient S/N to determine spectral characteristics of this star
\citep{Debosscher2013}. Given that both components of this binary
system are slow rotators showing large number of spectral lines in
their spectra, our denoising technique should be very robust in this
case. Figure~\ref{Figure12} illustrates two portions of the observed
composite spectrum of KIC\,11285625 along with the best fit LSD
model spectrum. A three-component average profile was computed for
each stellar component of this binary system, which means that we
solved for 2$\times$3=6 LSD profiles simultaneously. The line masks
for both stars were computed using the SynthV spectral synthesis
code \citep{Tsymbal1996} and assuming effective temperature and
surface gravity values of \te$^{1,2}$=6950/6400~K and
\logg$^{1,2}$=4.0/4.2~dex for the primary and secondary,
respectively \citep{Debosscher2013}. Given a low S/N of the spectrum
of the secondary component obtained in that work, effective
temperature of this star was estimated from the relative eclipse
depths. Similar to the case of KIC\,4749989 (cf. Section~4),
spectrum recovery for KIC\,11285625 is very good and a gain in S/N
is a factor of 10 compared to the original data. The high S/N LSD
spectra can further be used for spectral disentangling and ensure
high S/N for the resulting decomposed spectrum of the secondary. The
latter in turn can be used to determine the fundamental parameters
and chemical composition of the star.

In a similar way, our technique has been used to look for the
signatures of the secondary component in the (noisy) observed
spectra of a red giant star KIC\,05006817 \citep{Beck2013}. Using
the improved LSD technique, the authors could conclude that the star
is a single-lined binary, estimating the detection limit of the
spectral features of the secondary in the composite spectra to be of
the order of 1\% of the continuum.

Similar to the Section~3 dedicated to our improvements of the
original LSD method, we applied out technique to the binary spectrum
taken at the orbital phase of large RV separation. In practice, this
means that the individual mean profiles are computed from
non-overlapping RV grids, which provides their effective
deconvolution. A few additional tests that we conducted on synthetic
binary spectra characterized by a small RV separation of the two
stars reveal that the resulting individual mean profiles both show a
contribution of the primary and secondary components. In other
words, the LSD method suffers from degeneracy between spectral
contributions of the two stars when the average profiles are
computed on (largely) overlapping RV grids. Though the RVs derived
from such mean profiles can be associated with rather large
uncertainties, our line optimization method seems to work well also
at orbital phases of small RV separation, providing reasonably good
recovery of the original spectrum. Especially, this is the case when
the two stars show considerably different line broadening in their
spectra (have significantly different \vsini\ values). However, we
would expect our method to fail completely in eclipse phases, where
besides (nearly) zero RV separation of the stars, spectral lines of
one of the components can be remarkably distorted due to the
Rossiter-McLaughlin effect \citep{Rossiter1924,McLaughlin1924}.

\section{Discussion and Conclusions}\label{Section6}

The MOST, CoRoT, and $Kepler$ space-missions photometric data are of
unprecedented quality and led to a discovery of a large number of
intriguing objects (the stars hosting exoplanets, pulsating and
binary stars, etc.). For an efficient characterization of all those
diverse stellar objects, we still need ground-based support
observations in terms of high-resolution spectroscopy and/or
multicolour photometry. The majority of the objects are faint
implying that they either have to be observed with large telescopes
providing sufficiently high S/N for the desired analysis, or a
technique allowing to gain in S/N for the spectroscopic data
obtained with 2-m class telescopes is required. Thus, we dedicated
this paper to the development of such a technique, which is based on
the method of Least-Squares Deconvolution originally proposed by
\citet{Donati1997}.

We presented a generalization of the original LSD technique by
reconsidering its two fundamental assumptions on self-similarity of
all spectral lines and their linear addition to build the spectrum.
Following \citet{Kochukhov2010}, we implemented a multicomponent LSD
profile implying representation of the observed spectrum as a
superposition of several average profiles computed for different
line groups. The latter can be formed, e.g., according to the
strengths of individual lines, by subdividing the lines according to
the corresponding chemical elements (e.g., roAp or CP stars) and/or
stellar objects (the case of multiple stars), etc. We also
introduced a line strengths correction algorithm that minimizes the
effect of non-linear addition of spectral lines in a spectrum and
provides a good representation of the observations. The resulting
LSD model spectrum is of high S/N and contains all the information
from the original, observed spectrum. However, we stress that the
optimized individual line strengths hardly have their original
physical meaning, as the problem of non-linear addition of lines
with overlapping absorption coefficients requires a solution of
radiative transfer and cannot be solved by our simple line strength
optimization routine.

The method was first tested on simulated data of a fake stellar
object assuming three different values of S/N: infinity, 70, and 35.
We also checked the sensitivity of the technique to the chosen line
mask that in addition to the line positions contains an information
on their relative strengths. For calculation of the line masks, we
varied \te, \logg, and [M/H] in $\pm$500~K, $\pm$0.5~dex, and
$\pm$0.3~dex windows, respectively, and found that in all the cases,
our technique provides a good representation of the original
observed spectrum. The corresponding LSD model spectra were then
analysed to determine fundamental parameters by means of the
spectral synthesis method. The initially assumed parameter values
could always be efficiently reconstructed from the LSD model spectra
with the deviations from the original values of $\sim$50~K in \te,
$\sim$0.05~dex in both the metallicity and \logg, $\sim$0.5~\kms\ in
\vsini, and $\sim$0.18~\kms\ (the case of significantly wrong
metallicity, while the typical deviation is $\sim$0.06~\kms) in
microturbulent velocity being the worst.

The method was then applied to the observed spectra of Vega and a
$Kepler$ target, KIC04749989. Similar to the tests on simulated
data, the observed spectrum of Vega was successfully reproduced by
our LSD model and the derived fundamental parameters were found to
be in a good agreement with the recent findings from literature.
Since the method is certainly not aimed for application to the
bright stars like Vega for which a high-quality spectrum can be
obtained even with small telescopes, it was applied to a faint
target KIC04749989 for which a high-resolution but low S/N of
$\sim$60 spectrum existed. We found that such high noise level has a
large impact on the derived fundamental parameters when the original
observed spectrum is used for the analysis, but that a gain in S/N
by a factor of more than 10 overcomes this problem, and the
corresponding LSD model delivers appropriate for that star
atmospheric parameters.

A large number of faint pulsating stars discovered by, e.g.,
$Kepler$ space mission, nowadays requires ground-based support
(spectroscopic) observations to make the identification of the
individual pulsation modes possible for those stars. Thus, we tested
our method on the two pulsating stars, 20~CVn and HD\,189631, each
one oscillating in different types of modes, to check whether the
technique is also applicable to the intrinsically variable stars and
what kind of impact it might have on the pulsation content in the
spectrum, if any. We have found that in both cases the results are
very satisfactory and are in perfect agreement with those found in
the literature. In particular, we could show that 20~CVn is a
\DSct-type mono-periodic pulsator, most likely, oscillating in a
radial mode. These results were found to be consistent with the
findings by, e.g., \citet{Rodriguez1998}, \citet{Chadid2001}, and
\citet{Daszynska2003}. HD\,189631 was in turn proved to be a
multi-periodic \GD-type pulsator, with all the three dominant
frequencies being identified as $l$=1 sectoral modes. The findings
for the two dominant modes, f$_1$ and f$_2$, are compatible with the
recent results obtained by \citet{Maisonneuve2011} based on
high-resolution spectroscopic data. We also showed that the
identification presented by these authors for the third dominant
mode f$_4$ is in considerable disagreement with the observations,
which are in turn best represented by $l=1$ prograde sectoral mode.
We speculate that the three mode might show non-equidistant period
spacing of g-mode but more (theoretical) effort is required to prove
this.

We compared the LSD method with the two widely used smoothing
algorithms, running average and median filter. We found that the
latter are appropriate for constant stars with large projected
rotational velocities, so that lowering the originally high
($>$50\,000) resolving power of the spectrum has no impact on the
spectral characteristics derived from it. For slowly rotating and/or
intrinsically variable stars, the smoothing algorithms allow to
increase S/N of the spectra but deliver unreliable results. For
example, for sharp-lined stars, smoothing introduces artificial line
broadening which has an impact on all fundamental parameters, while
for the pulsating stars, it reduces the amplitude of or even
completely removes the pulsation signal. Our LSD method does not
reveal these problems and can be applied to both fast and slow
rotators as well as to the intrinsically variable stars.

Although the LSD method is expected to be the most effective for G-
to early A-type stars showing a lot of metal lines in their spectra,
it is not limited to these objects. It can also be applied to hotter
B-type stars given that large enough wavelength coverage is
provided. Though we did not test the method on these type of objects
in this paper, there are quite a few examples where the LSD
technique was successfully applied to hotter stars with the aim to
look for binary and/or spot signatures as well as to detect magnetic
fields
\citep[e.g.,][]{Petit2011,Kochukhov2011,Kochukhov_Sudnik2013,Tkachenko2012b,Aerts2013}.

Application to the spectrum of KIC\,11285625, a double-lined
spectroscopic binary, showed that the technique also allows
effectively denoise composite spectra of binary stars. In future
papers, we plan to continue development of the method and to
implement an algorithm of disentangling of binary spectra based on
the LSD profiles. Currently, it is impossible to do as our simple
line optimization algorithm suffers from strong degeneracy between
contributions of individual stellar components and delivers
unreliable disentangled spectra. However, at its present stage, the
method can also be effectively combined with the existing techniques
of disentangling of binary spectra. That is, one can apply the LSD
method to denoise composite spectra of a binary star and then to
disentangle the spectra of individual stellar components by means of
a disentangling technique in Fourier space
\citep{Hadrava2004,Ilijic2004}. KIC\,05006817, a red giant star in
an eccentric binary system, is one example of such simultaneous
application of the two methods \citep[for details, see][]{Beck2013}.
A complementary approach is to use the LSD profiles themselves for
accurate radial velocity measurements of spectroscopic binary and
triple systems \citet{Hareter2008}.

\begin{acknowledgements}
The research leading to these results received funding from the
European Research Council under the European Community's Seventh
Framework Programme (FP7/2007--2013)/ERC grant agreement
n$^\circ$227224 (PROSPERITY). OK is a Royal Swedish Academy of
Sciences Research Fellow, supported by the grants from Knut and
Alice Wallenberg Foundation and Swedish Research Council. Mode
identification results obtained with the software package {\sc
FAMIAS} developed in the framework of the FP6 European Coordination
Action HELAS (http://www.helas-eu.org/).
\end{acknowledgements}

{}

\end{document}